\documentclass[dvips]{acta}
\usepackage{supertabular,lscape,epsfig}
\usepackage{amssymb}
\usepackage{amsmath}
\mathchardef\mhyphen="2D
\usepackage{multirow}
\usepackage{colortbl}
\usepackage{placeins}%\Floatbarier
\DeclareSymbolFont{ppa}{OT1}{ppl}{m}{it}
\DeclareMathSymbol{\vv}{\mathalpha}{ppa}{'166}

\newfont{\hb}{rphvb at 10pt}%bezszeryfowe pó³grube
\newfont{\hbo}{rphvbo at 10pt}%bezszeryfowe pó³grube kursywa
\newfont{\bitt}{rptmbi at 12pt}%pó³gruba kursywa (tytu³ artyku³u)
\newfont{\bits}{rptmbi at 11pt}%pó³gruba kursywa (tytu³y rozdzia³ów)

\SetPages{149}{196}

\SetVol{66}{2016}

\begin{document}
\newcommand\pvalue{\mathop{p\mhyphen {\rm value}}}
%Zwarte naglowki, jeden wiersz, appendix
\newcommand{\TabApp}[2]{\begin{center}\parbox[t]{#1}{\centerline{
  {\bf Appendix}}
  \vskip2mm
  \centerline{\small {\spaceskip 2pt plus 1pt minus 1pt T a b l e}
  \refstepcounter{table}\thetable}
  \vskip2mm
  \centerline{\footnotesize #2}}
  \vskip3mm
\end{center}}

%Zwarte naglowki, jeden wiersz
\newcommand{\TabCapp}[2]{\begin{center}\parbox[t]{#1}{\centerline{
  \small {\spaceskip 2pt plus 1pt minus 1pt T a b l e}
  \refstepcounter{table}\thetable}
  \vskip2mm
  \centerline{\footnotesize #2}}
  \vskip3mm
\end{center}}

%Zwarte naglowki, dwa wiersze
\newcommand{\TTabCap}[3]{\begin{center}\parbox[t]{#1}{\centerline{
  \small {\spaceskip 2pt plus 1pt minus 1pt T a b l e}
  \refstepcounter{table}\thetable}
  \vskip2mm
  \centerline{\footnotesize #2}
  \centerline{\footnotesize #3}}
  \vskip1mm
\end{center}}

%Zwarte naglowki, jeden wiersz, appendix
\newcommand{\MakeTableApp}[4]{\begin{table}[p]\TabApp{#2}{#3}
  \begin{center} \TableFont \begin{tabular}{#1} #4 
  \end{tabular}\end{center}\end{table}}

%Zwarte naglowki, jeden wiersz
\newcommand{\MakeTableSepp}[4]{\begin{table}[p]\TabCapp{#2}{#3}
  \begin{center} \TableFont \begin{tabular}{#1} #4 
  \end{tabular}\end{center}\end{table}}

%Zwarte naglowki, jeden wiersz
\newcommand{\MakeTableee}[4]{\begin{table}[htb]\TabCapp{#2}{#3}
  \begin{center} \TableFont \begin{tabular}{#1} #4
  \end{tabular}\end{center}\end{table}}

%Zwarte naglowki, dwa wiersze
\newcommand{\MakeTablee}[5]{\begin{table}[htb]\TTabCap{#2}{#3}{#4}
  \begin{center} \TableFont \begin{tabular}{#1} #5 
  \end{tabular}\end{center}\end{table}}

%{\it Acta Astronomica Archive}
%\parskip=0pt \itemsep=1mm \setlength{\itemsep}{0.4mm}\setlength{\parindent}{-1em} \setlength{\itemindent}{-1em} - po \begin{itemize} - wszystko
%FWHM, PSF, S/N - proste, 
%MgII, H$\alpha$
%rms, rhs, sd - kursywa
%{\sc DAOPhot}
%{\sc Fnpeaks}
%{\sf files}
%Galactic wszystko (bulge, center, plane, disk, coordinates, latitudes...)
%Cepheids
%type~ Cepheids, Population~II Cepheids
%a.u.
%Polish National Science Centre
\newfont{\bb}{ptmbi8t at 12pt}
\newfont{\bbb}{cmbxti10}
\newfont{\bbbb}{cmbxti10 at 9pt}
\newcommand{\uprule}{\rule{0pt}{2.5ex}}
\newcommand{\douprule}{\rule[-2ex]{0pt}{4.5ex}}
\newcommand{\dorule}{\rule[-2ex]{0pt}{2ex}}

\begin{Titlepage}
  \Title{OGLE-ing the Magellanic System: Three-Dimensional Structure of the
    Clouds and the Bridge Using Classical Cepheids}

\Author{A.\,M.~~J~a~c~y~s~z~y~n~-~D~o~b~r~z~e~n~i~e~c~k~a$^1$,~~ 
D.\,M.~~S~k~o~w~r~o~n$^1$,~~
P.~~M~r~ó~z$^1$,\\
J.~~S~k~o~w~r~o~n$^1$,~~
I.~~S~o~s~z~y~ñ~s~k~i$^1$,~~
A.~~U~d~a~l~s~k~i$^1$,~~ 
P.~~P~i~e~t~r~u~k~o~w~i~c~z$^1$,\\
S.~~~K~o~z~³~o~w~s~k~i$^1$,~~ 
£.~~W~y~r~z~y~k~o~w~s~k~i$^1$,~~ 
R.~~P~o~l~e~s~k~i$^{1,2},$~~ 
M.~~P~a~w~l~a~k$^1$,\\
M.\,K.~~S~z~y~m~a~ñ~s~k~i$^1$~~ 
and~~K.~~U~l~a~c~z~y~k$^{1,3}$}
{$^1$Warsaw University Observatory, Al. Ujazdowskie 4, 00-478 Warszawa, Poland \\
e-mail: ajacyszyn@astrouw.edu.pl \\
$^2$Department of Astronomy, Ohio State University, 140 W. 18th Ave., Columbus~OH~43210, USA \\
$^3$Department of Physics, University of Warwick, Gibbet Hill Road, Coventry~CV4~7AL,~UK
}

\Received{March 1, 2016}
\end{Titlepage}

\Abstract{We analyzed a sample of 9418 fundamental-mode and first-overtone
  classical Cepheids from the OGLE-IV Collection of Classical Cepheids. The
  distance to each Cepheid was calculated using the period--luminosity
  relation for the Wesenheit magnitude, fitted to our data.

  The classical Cepheids in the LMC are situated mainly in the bar and in
  the northern arm. The eastern part of the LMC is closer to us and the
  plane fit to the whole LMC sample yields the inclination $i=24\zdot\arcd2
  \pm0\zdot\arcd7$ and position angle ${\rm P.A.}=151\zdot\arcd4
  \pm1\zdot\arcd7$. We redefined the LMC bar by extending it in the western
  direction and found no offset from the plane of the LMC contrary to
  previous studies. On the other hand, we found that the northern arm is
  offset from a plane by about $-0.5$~kpc, which was not observed
  before. The age distribution of the LMC Cepheids shows one maximum at
  about 100~Myr.

  We demonstrate that the SMC has a non-planar structure and can be
  described as an extended ellipsoid. We identified two large ellipsoidal
  off-axis structures in the SMC. The northern one is located closer to us
  and is younger, while the south-western is farther and older. The age
  distribution of the SMC Cepheids is bimodal with one maximum at 110~Myr,
  and another one at 220~Myr. Younger stars are located in the closer part
  of this galaxy while older ones are more distant.

  We classified nine Cepheids from our sample as Magellanic Bridge objects.
  These Cepheids show a large spread in three-dimensions although five of
  them form a connection between the Clouds. The closest one is closer than
  any of the LMC Cepheids, while the farthest one -- farther than any SMC
  Cepheid.  All but one Cepheids in the Magellanic Bridge are younger than
  300~Myr. The oldest one can be associated with the SMC Wing.}{Stars:
  fundamental parameters -- Cepheids -- Magellanic Clouds
  -- Galaxies: statistics -- Galaxies: structure}

\Section{Introduction}
\vspace*{11pt}
The Large Magellanic Cloud (LMC) and the Small Magellanic Cloud (SMC) are
one of our closest galaxies. What makes the LMC--SMC pair even more
interesting is that these galaxies have a common history. Their
interactions led to formation of a few intriguing structures: the
Magellanic Stream, the Leading Arm, and the Magellanic Bridge (Gardiner
\etal 1994, Gardiner and Noguchi 1996, Yoshizawa and Noguchi 2003, Connors
\etal 2006, R\r{u}\v{z}i\v{c}ka \etal 2009, 2010, Besla \etal 2010, 2012,
Diaz and Bekki 2011, 2012, Guglielmo \etal 2014). Together with the Magellanic
Clouds they constitute the Magellanic System.
\vspace*{3pt}

The Magellanic Stream is a 160\arcd\, long stream of gas that seems to be
trailing the Clouds' past orbit (Nidever \etal 2008, 2010). It has a double
nature in terms of morphology, velocity and metallicity (\eg Putman \etal
2003, Nidever \etal 2008, Fox \etal 2010, 2013, Richter \etal 2013). The
Leading Arm was formed together with the Stream (\eg Nidever \etal
2008). It comprises of four groups of High Velocity Cloud (Venzmer \etal
2012) and is interacting with matter in the Milky Way disk
(McClure-Griffiths \etal 2008). It is known to have a young stellar
component (Casetti-Dinescu \etal 2014).
\vspace*{3pt}

The Magellanic Bridge (MBR), a connection between the two Clouds, was known
as a gaseous feature since the work of Hindman \etal (1963). It is thought
to be formed after the last encounter of the LMC and SMC that took place
200--300~Myr ago (\eg Gardiner \etal 1994, Gardiner and Noguchi 1996,
R\r{u}\v{z}i\v{c}ka \etal 2010, Diaz and Bekki 2012, Besla \etal 2012). The
detailed analysis of neutral Hydrogen (HI) kinematics reveals that the
Magellanic Bridge is connected with the western part of the LMC disk (Indu
and Subramaniam 2015). Moreover, the velocity distribution suggests that
the MBR is being sheared. Numerical models predict that the Bridge should
have a stellar component (\eg Diaz and Bekki 2012, Besla \etal 2012,
Guglielmo \etal 2014), that should be an important tracer of interactions
between the LMC and SMC.
\vspace*{3pt}

Young stars in the area between the Clouds were observed by Shapley (1940).
Later, young stars were discovered farther from the SMC, in the direction
to the LMC (Irwin \etal 1985, Demers and Battinelli 1998, Harris 2007,
N\"oel \etal 2013, 2015). Finally, Skowron \etal (2014) showed that there
exists a continuous connection between the Clouds formed by a young stellar
population. Moreover, the Bridge also contains warm ionized gas (Barger
\etal 2013). Intermediate age stars were also observed in the MBR
(N\"oel \etal 2013, 2015), as well as candidates for an old stellar
population (Bagheri \etal 2013). Recent studies of stellar clusters and
associations suggest that these structures may be forming a tidal dwarf
galaxy (Bica \etal 2015) that had already been proposed by Bica and Schmitt
(1995). Such galaxies form from the gas pulled out of the interacting
galaxies and can have their own star formation (SF) processes (Ploeckinger
\etal 2014, 2015).

The interactions between the Magellanic Clouds have made a significant
impact on both galaxies. The knowledge of their structure brings relevant
implications for their common history as well as for other, more distant
galaxy systems. The Clouds are our closest interacting galaxies, thus can
be described as our ``local laboratory''. Their structure is also essential
for proper understanding of the nature of rare microlensing events detected
toward the Clouds and their interpretation either as self-lensing or due
to compact dark matter objects (\eg Wyrzykowski \etal 2011, Besla \etal
2013).

In the LMC younger and older stars have different spatial distributions
although the overall shape of the galaxy is roughly regular (\eg Cioni
\etal 2000, Bica \etal 2008, Joshi and Joshi 2014). Its disk is distorted,
elongated and asymmetrical and can be divided into inner and outer parts
with different inclination angles (van der Marel and Cioni 2001, van der
Marel 2001, Olsen and Salyk 2002, Nikolaev \etal 2004, Haschke \etal 2012a,
Subramanian and Subramaniam 2013). The eastern parts of the disk and the
halo are located closer to us because of the LMC's inclination toward the
SMC (van der Marel and Cioni 2001, Nikolaev \etal 2004, Persson \etal 2004,
Pejcha and Stanek 2009, Koerwer 2009, Subramanian and Subramaniam 2010,
Rubele \etal 2012, Haschke \etal 2012a, Subramanian and Subramaniam 2013,
van der Marel and Kallivayalil 2014, Deb and Singh 2014).

\hglue-7pt The LMC has an off-center bar that appears as an overdensity in
young and old stellar populations (Zhao and Evans 2000, Cioni \etal 2000,
van der Marel 2001, Nikolaev \etal 2004, Subramanian and Subramaniam 2013,
van der Marel and Kallivayalil 2014) as well as in the numerical models of
the off-center bar (Bekki 2009, Besla \etal 2012). The galaxy also has one
prominent spiral arm and maybe two or three irregular and not very
prominent arms (\eg Cioni \etal 2000, Nikolaev, \etal 2004, Bica \etal
2008, Moretti \etal 2014).  HI maps reveal four spiral-like structures
(Staveley-Smith \etal 2003) and the new ones have just been discovered
(Indu and Subramaniam 2015). Some of the LMC stars are kinematically
associated with these HI arms rather than with the disk (Olsen and Massey
2007).

The SMC is an elongated irregular galaxy with a central concentration where
young and old stars have slightly different distributions (\eg Cioni \etal
2000, Subramanian and Subramaniam 2012, Haschke \etal 2012b, Rubele \etal
2015). The SMC is known to have several substructures, of which the most
prominent is the Wing, that is a part of the galaxy that connects it with
the Magellanic Bridge (\eg Cioni \etal 2000, Nidever \etal 2011). Older
populations are more uniformly distributed while younger tend to
concentrate in the central parts and in the Wing. Moreover, the Wing also
comprises of many young stellar clusters (Piatti \etal 2015). Nidever \etal
(2013) showed that the optical depth in the eastern part of the SMC is two
times higher than in the western part, and the eastern part comprises of
two groups of stars with different mean distances. The SMC is rotated
toward the LMC and their closest parts on the sky are also the closest in
the sense of distance (Scowcroft \etal 2016).

The classical Cepheids (CCs) represent a young stellar population and play
an important role in structural studies of many extragalactic systems. In
the LMC and SMC they are of exceptional significance. Henrietta Leavitt had
discovered the famous Leavitt law studying the SMC Cepheids --
period--luminosity (P-L) relation -- Leavitt (1908).

Numerous studies of the LMC and SMC structure were based on the CCs.
Nikolaev \etal (2004) analyzed more than 2000 MACHO Cepheids in the LMC and
measured the viewing angles of this galaxy. They found that the results are
strongly dependent on the adopted center of the LMC, due to deviations from
the planar geometry. Moreover, they showed that the disk is warped, with
the bar being offset from the disk plane. A similar study was performed by
Persson \etal (2004) for 92 Cepheids observed in the near infrared
passbands. Later, Haschke \etal (2012ab) investigated almost 2000 Cepheids
from the OGLE-III data set. They constructed three-dimensional maps of the
Clouds by using individual reddening estimates and determining distances to
each Cepheid. They also detected mild twisting in the LMC disk and noticed
that the bar stands out as an overdensity.

Subramanian and Subramaniam (2015) fitted a plane to the SMC young stellar
``disk'' and found extra-planar features in front of and in the back of the
``disk''. The authors suggest that the former may be a tidal structure that
connects the SMC with MBR and the latter may be a stellar counterpart of
the Counter Bridge predicted by numerical models (Diaz and Bekki 2012). On
the other hand Scowcroft \etal (2016) showed that the SMC is extremely
elongated along the line of sight and they state that fitting a plane to
such structure is incorrect. The elongation of the SMC is consistent with
the significant optical depth values for this galaxy (\eg Nidever \etal
2013, Deb \etal 2015) and the numerical models predictions (Diaz and Bekki
2012).

The CCs were also used to study the star formation history (SFH) of the
Magellanic Clouds. Both galaxies have had an active SFH during the last
2~Gyr (Harris and Zaritsky 2009, Inno \etal 2015) and the age distribution
similarities between the LMC and SMC suggest that the galaxies must have
had common SF episodes (Harris and Zaritsky 2009, Indu and Subramaniam
2011, Inno \etal 2015, Subramanian and Subramaniam 2015, Joshi \etal 2016).

In this paper we present results of a three-dimensional analysis of the
Magellanic System using the OGLE Collection of Classical Cepheids recently
published by Soszyñski \etal (2015). The Collection is based on the OGLE-IV
data (Udalski \etal 2015), covering about 650 square degrees in this area.
Compared to the OGLE-III collection of Classical Cepheids, on which the
studies described above were based, the OGLE-IV Classical Cepheids
Collection includes the northern and southern parts of the LMC and extended
outskirts of the SMC. This is the first time that we see a full picture of
the Clouds with CCs from the OGLE project.

The sample completeness is over 99\%, which makes it the most complete and
least contaminated sample of CCs in the Magellanic Clouds and Bridge. Given
the vast OGLE-IV coverage of the Magellanic System, it is unlikely that
many additional CCs will be discovered in this region, making this the
ultimate collection of CCs in the Magellanic System.

The paper is organized as follows. In Section~2 we describe the OGLE-IV
data and OGLE Collection of Classical Cepheids. In Section~3 we present the
details of the analysis. Sections~4 to 6 contain results for the LMC, SMC,
and the Bridge, respectively. We discuss and summarize the results in
Sections~7 and~8.

\Section{Data}
\subsection{The OGLE Collection of Classical Cepheids}
The OGLE Collection of Classical Cepheids in the Magellanic System
(Soszyñ\-ski \etal 2015) contains 9535 objects of which 4620 are located
in the LMC and 4915 in the SMC OGLE-IV fields. Among those 5168 pulsate
solely in the fundamental mode (F), 3530 solely in the first-overtone
(1O), 117 oscillate only in the second-overtone (2O), 711 stars are
double-mode pulsators, and nine pulsate in three modes.

The collection is based on the {\it I}- and {\it V}-band photometry from
OGLE-IV (Udalski \etal 2015). The first step in variable star
classification was the visual inspection of candidates' light curves. The
selection of Cepheids was then based on the star's light curve shape, its
location in the P-L diagram, and the ratio of periods, if
multi-periodic. In some cases, the detailed inspection of the light curve
was repeated, taking other parameters of the star into account. The final
catalog contains CCs mean magnitudes in both bands, {\it I}-band amplitude,
pulsation periods, epochs of maximum light, and Fourier parameters derived
from the {\it I}-band light curves (Soszyñski \etal 2015).

\subsection{The Sample Selection}
For our analysis we chose CCs pulsating in the fundamental mode and the
first-overtone, including multi-mode pulsators, thus we excluded 117 stars
oscillating solely in the second overtone from our sample. We were left
with 9418 stars -- 4593 in the LMC and 4825 in the SMC. Among those, 32 CCs
(2 -- LMC and -- 30 SMC) are located in the genuine MBR fields, as defined
by OGLE-IV field names, \ie within RA $1\uph54\upm\lesssim\alpha
\lesssim4\uph06\upm$ (see green region in Fig.~19 of Udalski \etal 2015).

Next, we discarded Cepheids that did not have both {\it I}- and {\it
  V}-band magnitudes (50 objects from the LMC, 27 from the SMC and one from
MBR). Then, during the procedure of fitting the P-L relations to our sample
(see Section~3), we iteratively rejected Cepheids with the luminosity
deviating from the fit by more than $3\sigma$. This left us with 4222
Cepheids in the LMC, and 4663 in the SMC. We did not apply the fitting
procedure to the MBR Cepheids separately.

Soszyñski \etal (2015) state that at least five of the MBR CCs are truly
located in the MBR. We carefully inspected 31 objects from the genuine MBR
fields in terms of their location on the sky, distance from the observer
and from the Magellanic Clouds. Indeed, 22 of them ($\alpha\lesssim2\uph$)
are well correlated with the whole SMC sample, but nine are significantly
offset from both galaxies. We reclassify those as MBR stars. Thus the final
sample consists of 4222 Cepheids in the LMC, 4654 in the SMC and nine in
the MBR. The final sample numbers are summarized in Table~1.

\MakeTableee{c|ccccc}{6cm}{Classical Cepheid sample used in the analysis}
{\hline
\douprule
 & All & F & 1O & F1O\&1O2O & F1O2O\&1O2O3O\\
\hline
\uprule
LMC &   4222 &  2292 &  1589 &  337 & 4 \\ 
SMC &   4654 &  2646 &  1727 &  281 & 0 \\
MBR & ~~~~~9 &~~~~~4 &~~~~~4 &~~~~1 & ~0 \dorule\\
\hline
\douprule Total & 8885 & 4942 & 3320 & 619 & 4 \\
\hline
}

\Section{Data Analysis}
\subsection{Period-Luminosity Relation}
The first step in obtaining distances to Cepheids was to fit the P-L
relation to the LMC sample. In order to do this we first removed all the 1O
Cepheids with $\log P <-0.3$ (we express $P$ in days) from our sample. That
is because they may represent a different sample with different chemical
composition which is reflected in the P-L non-linearity near this value
(Soszyñski \etal 2008). Moreover, these stars are faintest, and most
affected by crowding and blending effects, hence have greater luminosity
uncertainty than the mean. For multi-mode pulsators we used the lowest
pulsation mode. For fitting we used the reddening-independent Wesenheit
magnitude (Madore 1976) for the {\it V}- and {\it I}-band photometry
defined as:
$$W_{I,V-I}=I-1.55\cdot(V-I).\eqno(1)$$
The coefficient 1.55 is calculated from a standard interstellar extinction
curve dependence of the {\it I}-band extinction on $E(V-I)$ reddening
(Schlegel, Finkbeiner and Davis 1998). We fitted a linear function in the
form:
$$W_{I,V-I}=a\cdot\textnormal{log}(P)+b\eqno(2)$$
using the least-squares method. In each iteration we rejected $3\sigma$
outliers until there were none. The majority of rejected outliers are due
to blending and crowding effects.

In the case of fundamental-mode CCs we divided the sample into two groups:
one with $\log P\leq0.4$, and one with $\log P>0.4$. A break in the P-L
relation at this value was already reported in the literature (\eg Bauer
\etal 1999, Udalski \etal 1999, Sharpee \etal 2002, Sandage \etal 2009,
Soszyñski \etal 2010). We also fitted the P-L relation to the {\it I}- and
{\it V}-band magnitudes (without correcting for extinction). The same
procedure was repeated for the SMC. Results are shown in Table~2 and in
Fig.~1.

\begin{figure}[p]
\vglue-5pt
\includegraphics[width=12cm]{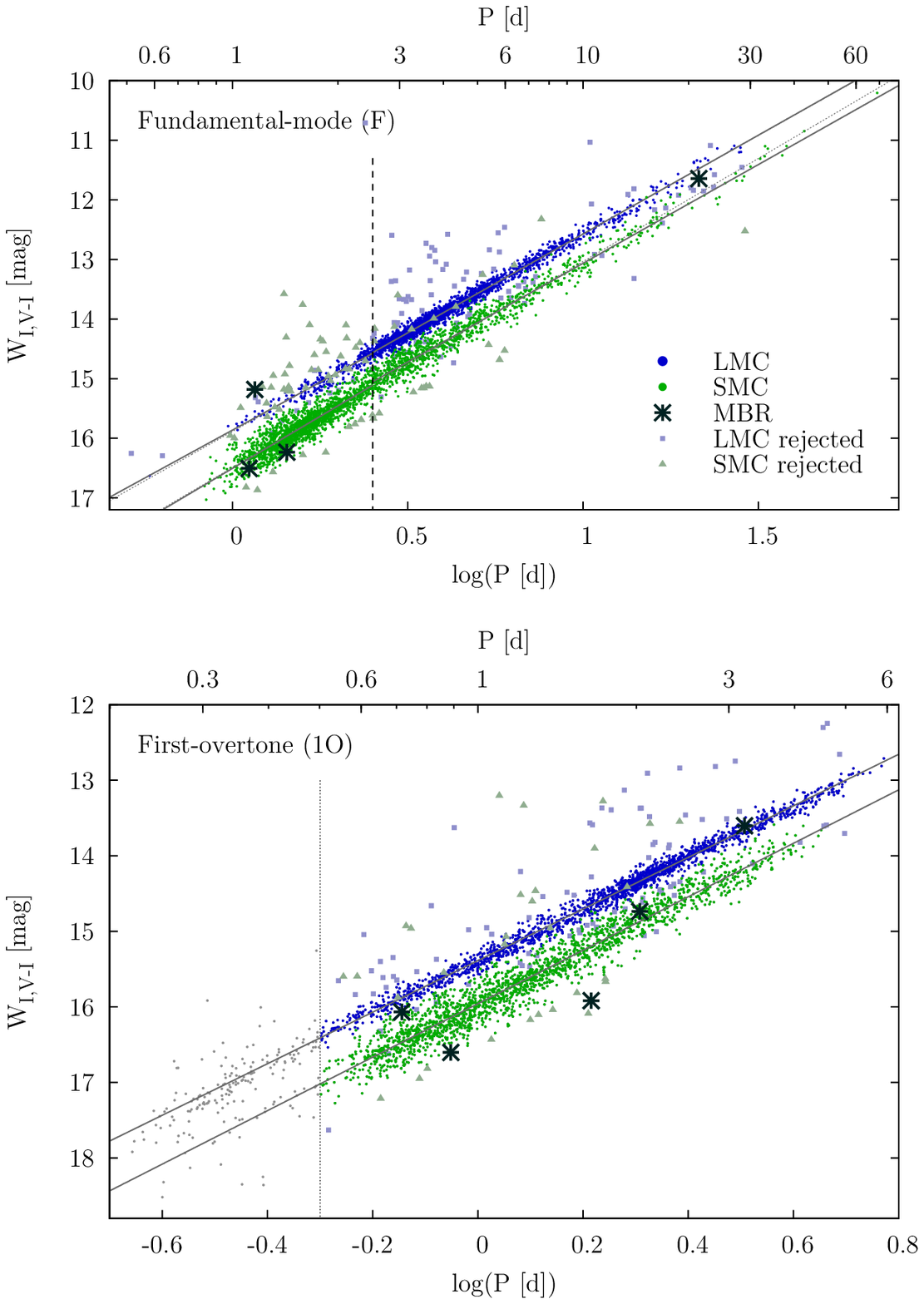}
\FigCap{P-L relations for the Wesenheit magnitude for fundamental-mode
  ({\it top panel}) and first-overtone ({\it bottom panel}) CCs in the LMC
  (blue dots) and in the SMC (green dots). The MBR Cepheids are marked with
  large stars. Gray points represent stars rejected during the iterative
  $3\sigma$ clipping when fitting the P-L relations. {\it Top panel:} Solid
  lines represent separate fits for two sets of F Cepheids divided at $\log
  P=0.4$. The dotted lines show fits for the whole F sample in the LMC and
  SMC. The dashed vertical line represents the period at which the P-L
  relation breaks. {\it Bottom panel:} The solid lines show fits for 1O
  CCs. Small gray dots represent 1O Cepheids with $\log
  P<-0.3$ that were removed from our sample, as marked by the dotted
  vertical line.}
\end{figure}
\definecolor{lightgray}{gray}{0.60}
\renewcommand{\arraystretch}{1.07}
\MakeTableee{c@{\hspace{5pt}}c@{\hspace{5pt}}c@{\hspace{5pt}}!{\color{black}\vrule}c!{\color{black}\vrule}c!{\color{black}\vrule}c!{\color{black}\vrule}r@{.}l!{\color{black}\vrule}c!{\color{black}\vrule}c}{12.5cm}{P-L relations for CCs in the Magellanic Clouds}
{
\hline
\noalign{\vskip5pt}
\multicolumn{3}{l}{P-L for Wesenheit magnitude} & \multicolumn{7}{c}{$W_{I,V-I}=a\cdot\log P+b$}\\ 
\noalign{\vskip5pt}
\hline 
\noalign{\vskip2pt}
Galaxy & P. mode & $\log P$ & $a$ & $b$ [mag] & $\sigma$ [mag] & \multicolumn{2}{c|}{$\chi^2/{\rm dof}$} & $N_{\rm inc}$ & $N_{\rm rej}$ \\ 
\noalign{\vskip2pt}
\hline
\multirow{4}{*}{LMC} & \multirow{3}{*}{F} & $\leq0.4$ & $-3.216\pm0.033$ & $15.864\pm0.010$ & 0.103 & 2 & 991 & 284 & 6 \\
&    & $>0.4$ & $-3.317\pm0.007$ & $15.890\pm0.005$ & 0.075 & 1&568 & 2103 & 87 \\
&    & all    & $-3.313\pm0.006$ & $15.888\pm0.004$ & 0.078 & 1&686 & 2382 & 98 \\ 
\arrayrulecolor{lightgray}\cline{2-10}
& 1O & all    & $-3.414\pm0.007$ & $15.388\pm0.002$ & 0.079 & 1&714 & 1931 & 84 \\ 
\arrayrulecolor{black}\hline
\multirow{4}{*}{SMC} & \multirow{3}{*}{F} & $\leq0.4$ & $-3.488\pm0.015$ & $16.507\pm0.004$ & 0.157 & 6&920 & 1746 & 43 \\
&    & $>0.4$ & $-3.315\pm0.009$ & $16.379\pm0.006$ & 0.144 & 5&811 & 957  & 30 \\
&    & all    & $-3.458\pm0.005$ & $16.492\pm0.002$ & 0.155 & 6&746 & 2708 & 68 \\ 
\arrayrulecolor{lightgray}\cline{2-10}
& 1O & all    & $-3.540\pm0.007$ & $15.959\pm0.002$ & 0.170 & 8&083 & 2010 & 30 \\ 
\arrayrulecolor{black}\hline
\noalign{\vskip5pt}
\multicolumn{3}{l}{P-L for {\it I}-band magnitude} & \multicolumn{7}{c}{$I=a\cdot\log(P)+b$}\\ 
\noalign{\vskip5pt}
\hline
\noalign{\vskip2pt}
Galaxy & P. mode & $\log P$ & $a$ & $b$ [mag]& $\sigma$ [mag] & \multicolumn{2}{c|}{$\chi^2/{\rm dof}$} & $N_{\rm inc}$ & $N_{\rm rej}$ \\ 
\noalign{\vskip2pt}
\hline
\multirow{4}{*}{LMC} & \multirow{3}{*}{F} & $\leq0.4$ & $-3.036\pm0.032$ & $16.865\pm0.010$ & 0.140 & 5&499 & 279 & 11 \\
&    & $>0.4$ & $-2.894\pm0.007$ & $16.810\pm0.005$ & 0.147 & 6&015 & 2093 & 97  \\
&    & all    & $-2.911\pm0.006$ & $16.822\pm0.004$ & 0.146 & 5&959 & 2372 & 108 \\ 
\arrayrulecolor{lightgray}\cline{2-10}
& 1O & all    & $-3.240\pm0.006$ & $16.356\pm0.002$ & 0.159 & 7&065 & 1950 & 65  \\ 
\arrayrulecolor{black}\hline
\multirow{4}{*}{SMC} & \multirow{3}{*}{F} & $\leq0.4$ & $-3.147\pm0.015$ & $17.420\pm0.004$ & 0.208 & 12&104 & 1756 & 33 \\
&    & $>0.4$ & $-2.912\pm0.009$ & $17.241\pm0.006$ & 0.222 & 13&815 & 976 & 11 \\
&    & all    & $-3.113\pm0.005$ & $17.401\pm0.002$ & 0.216 & 13&064 & 2734 & 42 \\ 
\arrayrulecolor{lightgray}\cline{2-10}
& 1O & all    & $-3.278\pm0.007$ & $16.813\pm0.002$ & 0.223 & 13&916 & 2007
& 33 \\ 
\arrayrulecolor{black}\hline
\noalign{\vskip5pt}
\multicolumn{3}{l}{P-L for {\it V}-band magnitude} & \multicolumn{7}{c}{$V=a\cdot\log(P)+b$}\\ 
\noalign{\vskip5pt}
\hline
\noalign{\vskip2pt}
Galaxy & P. mode & $\log P$ & $a$ & $b$ [mag]& $\sigma$ [mag]& \multicolumn{2}{c|}{$\chi^2/{\rm dof}$} & $N_{\rm inc}$ & $N_{\rm rej}$ \\ 
\noalign{\vskip2pt}
\hline
\multirow{4}{*}{LMC} & \multirow{3}{*}{F} & $\leq0.4$ & $-2.964\pm0.032$ & $17.526\pm0.010$ & 0.190 & 10&142 & 280 & 10 \\
&    & $>0.4$ & $-2.629\pm0.007$ & $17.399\pm0.005$ & 0.211 & 12&412 & 2090 & 100 \\
&    & all    & $-2.672\pm0.006$ & $17.429\pm0.004$ & 0.207 & 11&986 & 2365 & 115 \\ 
\arrayrulecolor{lightgray}\cline{2-10}
& 1O & all    & $-3.133\pm0.006$ & $16.975\pm0.002$ & 0.223 & 13&983 & 1946 & 69  \\ 
\arrayrulecolor{black}\hline
\multirow{4}{*}{SMC} & \multirow{3}{*}{F} & $\leq0.4$ & $-2.914\pm0.015$ & $18.001\pm0.004$ & 0.254 & 18&003 & 1758 & 31 \\
&    & $>0.4$ & $-2.648\pm0.009$ & $17.792\pm0.006$ & 0.283 & 22&469 & 978  &  9 \\
&    & all & $-2.901\pm0.005$    & $17.984\pm0.002$ & 0.266 & 19&846 & 2734 & 42 \\ 
\arrayrulecolor{lightgray}\cline{2-10}
& 1O & all & $-3.122\pm0.007$    & $17.361\pm0.002$ & 0.273 & 20&912 & 2004 & 36 \\ 
\arrayrulecolor{black}\hline
\noalign{\vskip5pt}
\multicolumn{10}{p{12cm}}{$N_{\rm inc}$ is the number of objects included in the final fit, while $N_{\rm rej}$ is the number of rejected objects.}
}

The most accurate fits are obtained for the Wesenheit magnitude for the LMC
Cepheids. They show the smallest scatter of only 0.08 mag. This is why we
decided to use these relations for distance determinations in further
analysis. In the case of the SMC, large values of $\chi^2/{\rm dof}$ are a
result of this galaxy's elongation almost along the line of sight --
significant distance differences between the Cepheids account for the
scatter in magnitudes.

The slopes of the P-L for the Wesenheit index for F Cepheids with $\log
P>0.4$ are identical for the LMC and SMC within $1\sigma$ errors, as
expected (Ngeow \etal 2015). We cannot compare slopes for $\log P<0.4$ for
two reasons. First, the LMC sample is much less numerous than the SMC
sample and so the comparison would be biased (Udalski \etal 1999). Second,
the SMC may simply have a different value of the slope because of its
different environment and Cepheids with shorter periods may have different
chemical composition (Bauer \etal 1999, Soszyñski \etal 2010). When
calculating the distances we assume that the SMC $\log P<0.4$ slope is
identical as for the LMC.

\subsection{Distances}
In order to obtain both LMC and SMC Cepheid distances we used the mean
distance to the LMC measured by Pietrzyñski \etal (2013) from
eclipsing-binaries, $d_{\rm LMC}=49.97\pm0.19~{\rm (statistical)}\pm1.11~
{\rm (systematic)}$~kpc. With 2.2\% error it is the most accurate
measurement of the mean LMC distance up to date.

For each object we calculated the reference magnitude $W_{\rm ref}$, \ie
the Wesenheit magnitude on the fitted P-L line (for the LMC) corresponding
to its period $P$:
$$W_{\rm ref}=a_{{\rm LMC}}\cdot\log(P)+b_{{\rm LMC}}.\eqno(3)$$

We used $a$ and $b$ coefficients from Table~2, in the case of
fundamental-mode Cepheids separately for $\log P\leq0.4$ and $>0.4$. So the
relative distance modulus is:
$$\delta\mu=W_{I,V-I}-W_{\rm ref}.\eqno(4)$$
And then the absolute distance simply:
$$d=d_{{\rm LMC}}\cdot10^{\frac{\delta\mu}{5}}.\eqno(5)$$

\begin{figure}[htb]
\includegraphics[width=12.5cm]{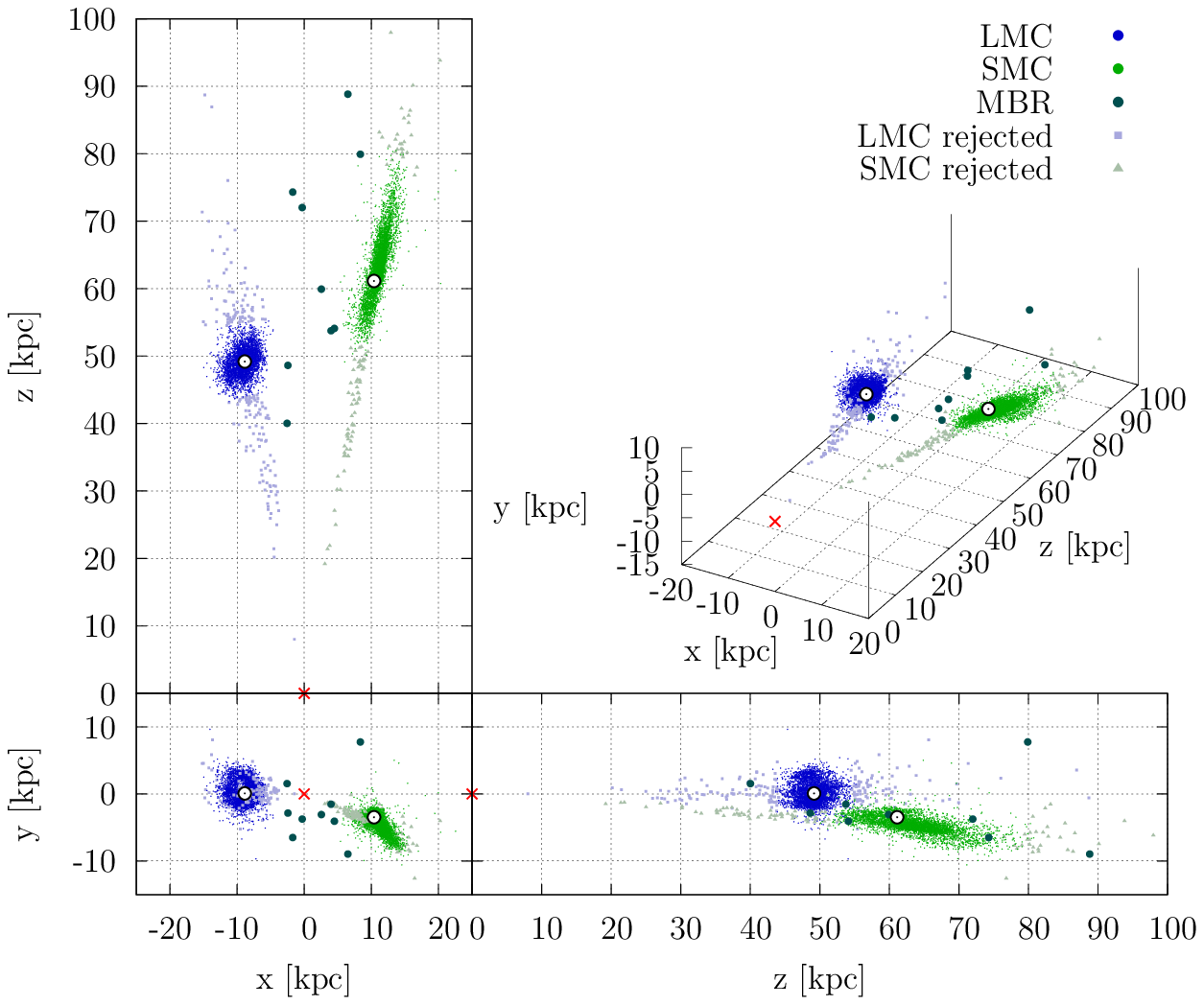}
\FigCap{Three-dimensional map showing the Magellanic System in Cartesian
  coordinates with the {\it z} axis pointing toward $\alpha_{\rm
  cen}=3\uph20\upm$, $\delta_{\rm cen}=-72\arcd$. Blue dots represent the
  LMC, green dots SMC, and the large dark teal dots -- MBR. Gray points
  show the $3\sigma$ outliers rejected in the P-L fitting procedure (see
  Fig.~1 for comparison). Red cross marks the observer's location. White
  circles mark the LMC (Pietrzyñski \etal 2013, van der Marel and
  Kallivayalil 2014) and SMC (Graczyk \etal 2014, Stanimiroviæ \etal
  2004) centers.}
\end{figure}
Fig.~2 shows three-dimensional maps of the Magellanic System in the
Cartesian space. Blue dots mark the LMC Cepheids, green dots SMC, and large
dark teal dots show the Magellanic Bridge sample. Gray points mark the
$3\sigma$ outliers rejected in the procedure of P-L fitting (see Fig.~1 for
comparison). There is a distinct spread in the Cepheid distances along the
line of sight that is mostly, but not entirely physical, and a part of it
is due to errors in distance calculation. The errors are typically
1.2--1.6~kpc (median $\approx3\%$ relative) for the LMC and 1.4--2.3~kpc
(median $\approx3\%$ relative) for the SMC. When calculating the
uncertainties we used the error of zero points of the OGLE-IV photometry
which is $\sigma_{I,V}=0.02$~mag and the uncertainties of the P-L fit which
are shown in Table~2. We intentionally omitted the uncertainty of the LMC
distance measurement because it would only increase Cepheid distance
uncertainties without affecting the geometry. While the photometry error
itself is not large $\sigma_{I,V}=0.02$~mag, it translates at the LMC
distance to $\sigma_{d,I,V}=0.46$~kpc and $\sigma_{d,W}=0.65$~kpc and this
is the ``natural spread'' of the method. There is also a possibility, that
even though we are using the reddening-free Wesenheit index, the
differential and variable extinction within the LMC/SMC may add up to
the error in Cepheid distances.

We have analyzed how much the adopted reddening law influences the distance
uncertainties. For a Wesenheit index with a different coefficient:
$$W_{I,V-I}=I-1.44\cdot(V-I)\eqno(6)$$
(Udalski 2003) we obtained slightly smaller uncertainties. In the case of
the LMC the median distance uncertainty was about 1.38~kpc (2.8\% relative)
when using a coefficient of 1.55, and 1.31~kpc (2.6\% relative) when using
1.44. In the case of the SMC the numbers are: 1.79~kpc (2.8\% relative) for
1.55, and 1.70~kpc (2.6\% relative) for 1.44. We see that the choice of
the reddening law coefficient does not influence the distance uncertainties
in a significant way.

\subsection{Coordinate Transformations}
In this study we visualize the results with two types of maps. The first one is
a two-dimensional sky map in a Hammer equal-area projection. The projection
is rotated so that the {\it z} axis is pointing toward $\alpha_{\rm cen}=3\uph20\upm$,
$\delta_{\rm cen}=-72\arcd$. For each Cepheid,
$x_{\rm Hammer}$ and $y_{\rm Hammer}$ are calculated from:

\setcounter{equation}{6}
\begin{eqnarray}
\alpha_{\rm b} &=&\alpha+\left(\frac{\pi}{2}-\alpha_{\rm cen}\right),\\
l              &=&\arctan\left(\frac{\sin(\alpha_{{\rm b}})\cos(\delta_{\rm cen})+\tan(\delta)\sin(\delta_{\rm cen})}{\cos(\alpha_{{\rm b}})}\right),\ l \in [-\pi;\pi],\\
\beta          &=&\arcsin(\sin(\delta)\cos(\delta_{\rm cen})-\cos(\delta)\sin(\delta_{\rm cen})\sin(\alpha_{{\rm b}})),\\
x_{\rm Hammer} &=&-\frac{2\sqrt{2}\cdot\cos(\beta)\sin\left(\frac{l}{2}\right)}{\sqrt{1+\cos(\beta)\cos\left(\frac{l}{2}\right)}},\\
y_{\rm Hammer} &=&\frac{\sqrt{2}\cdot\sin(\beta)}{\sqrt{1+\cos(\beta)\cos\left(\frac{l}{2}\right)}}.
\end{eqnarray}
Fig.~3 shows the Hammer projection of the Magellanic System where the
Cepheid distances are color-coded. The LMC is on the left, with a clearly
visible bar and a northern arm, while the SMC is on the right. The Magellanic
Bridge Cepheids between the two galaxies are marked with larger dots. Here
we can clearly see the distance differences between the two galaxies. The
bottom panels show close-ups of each of the Clouds. When looking at the LMC
(left) we can clearly see the inclination of this galaxy -- the western
side of the LMC (the right side of the map) lies farther from us than the
eastern side.  In fact, it is rotated in the direction of the SMC. The
right panel shows the SMC close-up. The large spread in Cepheid distances
reflects the galaxy's significant elongation (see Fig.~2 for comparison).

\begin{figure}[htb]
\includegraphics[width=12.5cm]{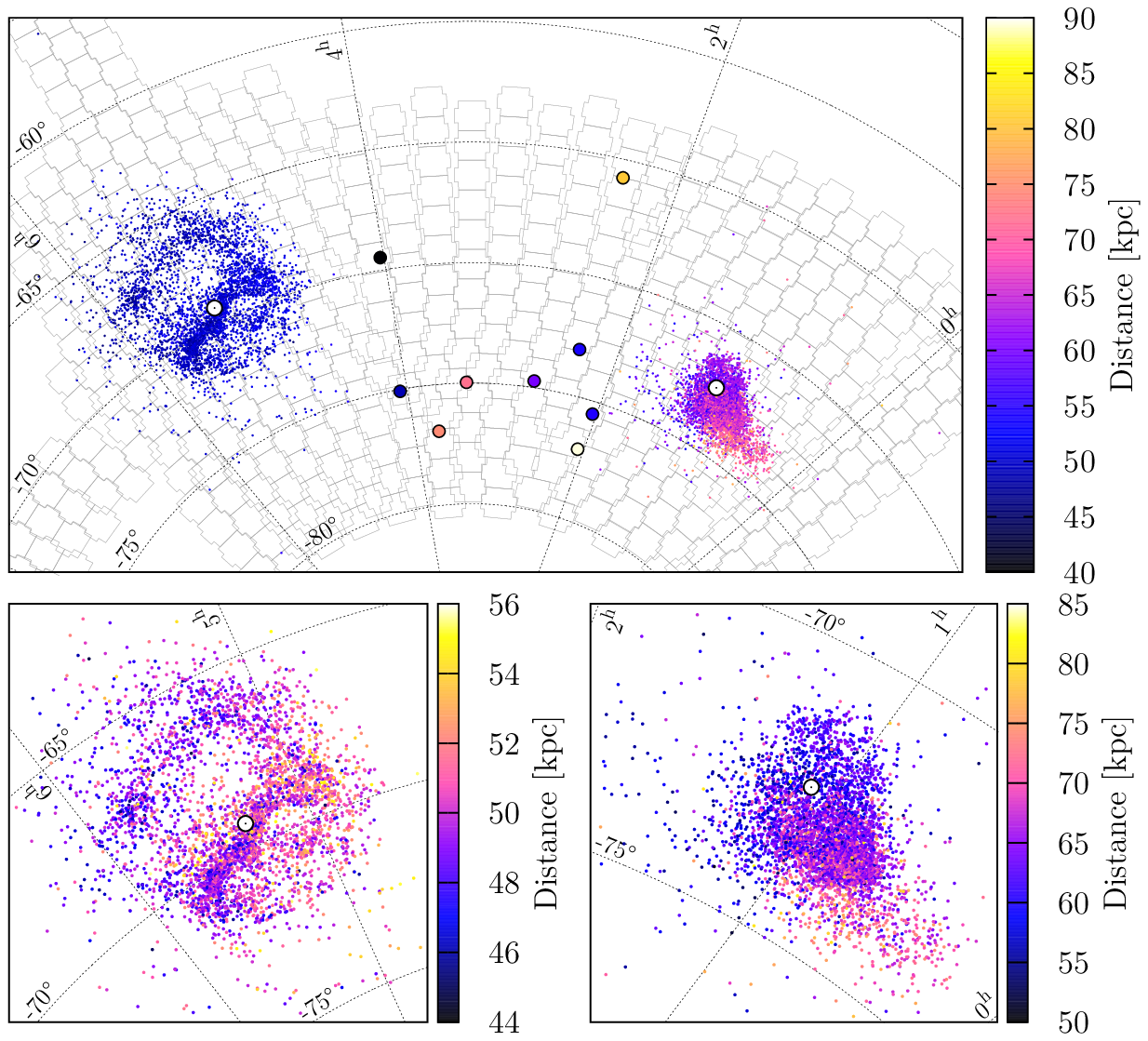}
\FigCap{Three-dimensional map of CCs in the Magellanic System in a Hammer
  projection with the {\it z} axis pointing toward $\alpha_{\rm
    cen}=3\uph20\upm$, $\delta_{\rm cen}=-72\arcd$. Cepheid distances are
  color-coded. {\it Upper panel:} MBR Cepheids are marked with large
  dots. Gray contours represent OGLE-IV fields in the Magellanic System.
  {\it Lower left panel:} Close-up on the LMC. {\it Lower right panel:}
  Close-up on the SMC.  Note the change in color range. White circles mark
  the LMC (van der Marel and Kallivayalil 2014) and SMC (Stanimiroviæ \etal
  2004) centers.}
\end{figure}

The second type of maps used in this study are the three-dimensional
Cartesian space ({\it x}, {\it y}, {\it z}) projections with different
viewing angles. In this transformation the observer is always in $(0,0,0)$
while the {\it z} axis is pointing toward different equatorial
coordinates: $\alpha_{\rm cen}$ and $\delta_{\rm cen}$. The transformation
equations were taken from van der Marel and Cioni (2001) and Weinberg and
Nikolaev (2001):

\begin{eqnarray}
        x &=& -d\cdot\cos(\delta)\sin(\alpha-\alpha_{\rm cen}),\\
        y &=&~~d\cdot(\sin(\delta)\cos(\delta_{\rm cen})-\cos(\delta)\sin(\delta_{\rm cen})\cos(\alpha-\alpha_{\rm cen})),\\
        z &=&~~d\cdot(\cos(\delta)\cos(\delta_{\rm cen})\cos(\alpha-\alpha_{\rm cen})+\sin(\delta)\sin(\delta_{\rm cen})),
\end{eqnarray}
where $d$ is the calculated distance to each Cepheid and $\alpha_{\rm
cen}$, $\delta_{\rm cen}$ are the map center coordinates.  Maps showing
only the LMC or only SMC are rotated so that the {\it z} axes are pointing
toward their dynamical centers.  For the LMC we adopt $\alpha_{\rm
LMC-cen}=5\uph20\upm12\ups$, $\delta_{\rm LMC-cen}=-69\arcd18\arcm$,
which is for the whole population with a correction for young stars proper
motions (van der Marel and Kallivayalil 2014).  For the SMC we use
$\alpha_{\rm SMC-cen}=1\uph05\upm$, $\delta_{\rm
SMC-cen}=-72\arcd25\arcm12\arcs$ (Stanimiroviæ \etal 2004).  We decided
to use the dynamical centers of these galaxies because we think they are
the most reliable. The same centers were used to calculate Magellanic
Clouds' proper motions (see Kallivayalil \etal 2006ab, 2013 and van der
Marel and Kallivayalil 2014).

The uncertainties of the Cartesian coordinates include the OGLE astrometric
uncertainty which is $\sigma_{\alpha,\delta}=0\zdot\arcs2$. Every
coordinate is also dependent on the distance, so the uncertainties of {\it
x}, {\it y} and {\it z} include the distance uncertainty. Their values
are in the following ranges: $0.4~{\rm kpc}<\sigma_x<1.3~{\rm kpc}$,
$0.6~{\rm kpc}<\sigma_y<1.3~{\rm kpc}$, and $1.3~{\rm kpc}<
\sigma_z<2.4~{\rm kpc}$.

The most important parameters of the CCs sample analyzed in
this publication are available online from the OGLE website:

\centerline{\it $$http://ogle.astrouw.edu.pl$$}

Table~3 presents the first few lines of the data file.

\renewcommand{\TableFont}{\scriptsize}
\MakeTableee{c@{\hspace{3pt}}c@{\hspace{3pt}}c@{\hspace{7pt}}c@{\hspace{7pt}}c@{\hspace{7pt}}c@{\hspace{3pt}}c}{12cm}{Classical Cepheids in the Magellanic System}
{
\hline
\noalign{\vskip3pt}
\multicolumn{7}{c}{Columns 1-7} \\ \hline
\noalign{\vskip3pt}
Location & OCVS Id & P. mode & P$^{(a)}$ [d] & {\it I} [mag] & {\it V} [mag]
& $W_{I,V-I}$ [mag] \\ 
\noalign{\vskip5pt}
\hline
\noalign{\vskip3pt}
LMC & OGLE-LMC-CEP-0004 &       1O &  2.2296385 & 15.123 & 15.690 & 14.244 \\ 
LMC & OGLE-LMC-CEP-0005 &        F &  5.6119491 & 14.651 & 15.425 & 13.451 \\ 
LMC & OGLE-LMC-CEP-0006 &       1O &  3.2947501 & 14.707 & 15.366 & 13.686 \\
LMC & OGLE-LMC-CEP-0007 &       1O &  0.7090827 & 16.955 & 17.561 & 16.016 \\ 
LMC & OGLE-LMC-CEP-0008 &    1O/2O &  0.9728732 & 16.337 & 16.921 & 15.432 \\ 
$\vdots$ & $\vdots$ & $\vdots$ & $\vdots$ & $\vdots$ & $\vdots$ & $\vdots$ \\ \cline{1-7}
\noalign{\vskip3pt}
\multicolumn{7}{c}{Columns 8-14} \\ \hline
\noalign{\vskip5pt}
RA & Dec & d [kpc] & $x^{(b)}$ [kpc] & $y^{(b)}$ [kpc] & $z^{(b)}$ [kpc] &
Age$^{(c)}$ [Myr] \\ 
\noalign{\vskip3pt}
\hline
\noalign{\vskip3pt}
$04\uph35\upm20\zdot^{\rm s}16$ & $-69\arcd48\arcm07\zdot\arcs7$ & $51.03 \pm 1.40$ &  $-5.69 \pm 0.43$ &   $1.06 \pm 0.85$ &  $50.70 \pm 1.50$ & $102 \pm 19$ \\
$04\uph35\upm31\zdot^{\rm s}52$ & $-69\arcd44\arcm05\zdot\arcs8$ & $51.05 \pm 1.41$ &  $-5.72 \pm 0.43$ &   $1.11 \pm 0.85$ &  $50.72 \pm 1.50$ &  $66 \pm 15$ \\
$04\uph35\upm42\zdot^{\rm s}16$ & $-69\arcd43\arcm29\zdot\arcs1$ & $51.51 \pm 1.42$ &  $-5.79 \pm 0.43$ &   $1.13 \pm 0.86$ &  $51.18 \pm 1.51$ &  $75 \pm 14$ \\
$04\uph36\upm30\zdot^{\rm s}06$ & $-68\arcd37\arcm35\zdot\arcs7$ & $52.77 \pm 1.45$ &  $-6.30 \pm 0.46$ &   $2.10 \pm 0.88$ &  $52.35 \pm 1.55$ & $256 \pm 47$ \\
$04\uph36\upm33\zdot^{\rm s}08$ & $-69\arcd18\arcm43\zdot\arcs6$ & $50.05 \pm 1.38$ &  $-5.80 \pm 0.43$ &   $1.42 \pm 0.84$ &  $49.69 \pm 1.47$ & $199 \pm 36$ \\
$\vdots$ & $\vdots$ & $\vdots$ & $\vdots$ & $\vdots$ & $\vdots$ & $\vdots$\\ 
\cline{1-7}
\noalign{\vskip3pt}
\multicolumn{7}{p{12cm}}{The electronic version of the whole sample used
in this study is available online from the OGLE website. {\it (a)} 
For multi-mode Cepheids the longest period is provided. {\it (b)} 
The cartesian {\it x}, {\it y} and {\it z} coordinates. 
{\it (c)}~The ages were calculated using PA relations from Bono \etal (2005).}
}

\subsection{Model and Plane Fitting}
In the next step we attempt to characterize the LMC Cepheids in three
dimensions. Here we use a Cartesian coordinate system with the origin in
the LMC center and {\it z} axis oriented toward the observer.
\begin{eqnarray}
 	x=d\tilde{x}(\alpha,\delta)&{=}&-d\cdot\cos(\delta)\sin(\alpha-\alpha_{{\rm LMC-cen}}),\\
	y=d\tilde{y}(\alpha,\delta)&{=}&d\cdot(\sin(\delta)\cos(\delta_{\rm LMC-cen})-\nonumber\\
                                   & &-\cos(\delta)\sin(\delta_{\rm LMC-cen})\cos(\alpha-\alpha_{\rm LMC-cen}),\\
z{=}d_{\rm LMC}{-}d\tilde{z}(\alpha,\delta)&{=}&d_{\rm LMC}{-}d{\cdot}(\cos(\delta)\cos(\delta_{\rm LMC{-}cen})\cos(\alpha{-}\alpha_{\rm LMC{-}cen}){+}\nonumber\\
                                   & &+\sin(\delta)\sin(\delta_{\rm LMC-cen})).
\end{eqnarray}
Structural parameters of the LMC disk (inclination, position angle) can be
inferred from a plane fit to the data:
$$z=ax+by+c.\eqno(18)$$

The coefficient $c$ quantifies the shift of the best-fit plane from the
adopted LMC center. The remaining two parameters can be transformed to the
disk inclination $i$ and position angle P.A.:
\setcounter{equation}{18}
\begin{eqnarray}
	 i &=&\arccos\left(1/\sqrt{a^2+b^2+1}\right),\\
{\rm P.A.} &=&\arctan\left(-\frac{a}{b}\right)+\frac{\pi}{2}{\rm sgn}{({\it b})}.
\end{eqnarray}

A simple linear least-squares fit to the plane equation can be biased,
because the uncertainties of all coordinates ({\it x}, {\it y}, {\it z})
are not negligible, since they all contain distance measurement
error. Hence, we propose a parametrization in which a line joining the
observer and the {\it i}-th Cepheid intersects the fitted plane at a
distance:
$$d_{\rm model}(\alpha_i,\delta_i;a,b,c)=\frac{d_{\rm LMC}-c}{\tilde{z}(\alpha_i,\delta_i)+a\tilde{x}(\alpha_i,\delta_i)+b\tilde{y}(\alpha_i,\delta_i)}\eqno(21)$$
or a distance modulus:
$$\mu_{\rm model}(\alpha_i,\delta_i;a,b,c)=5\log(d_{\rm model}(\alpha_i,\delta_i;a,b,c))+10\eqno(22)$$
if $d_{{\rm model}}$ is expressed in kpc. We minimize the sum:
$$\chi^2(a,b,c)=\sum_{i}\left(\frac{\mu_i-\mu_{\rm model}(\alpha_i,\delta_i;a,b,c)}{\sigma_{\mu,i}}\right)^2\eqno(23)$$
using the Nelder-Mead algorithm (Nelder and Mead 1965). The adopted
uncertainties $\sigma_{\mu,i}$ include OGLE photometry uncertainties
($\sigma_{I,V}=0.02$~mag) and the uncertainties of the P-L fit given in
Table~2. The fitting procedure is iterative and after each step $3\sigma$
outliers are rejected. The typical deviation from the best-fit plane
(1.5~kpc) is constrained by the accuracy of the P-L relation and the
``natural spread'' of the method of calculating distances as described
above (0.65~kpc). We checked that the influence of the choice of $d_{\rm
LMC}$ and $(\alpha_{{\rm LMC-cen}}, \delta_{{\rm LMC-cen}})$ on the
best-fit parameters is negligible.

\Section{The Large Magellanic Cloud}
\subsection{Three-Dimensional Structure}
Previous studies of the LMC CCs based on the OGLE-III data (\cf Fig.~1 from
Haschke \etal 2012a) did not include the northern and southern parts of the
galaxy. This is the first time that we see a full picture of the LMC with
the OGLE Cepheids.

Fig.~3 shows that the disk of the LMC is inclined and rotated in the
direction of its smaller neighbor, the SMC. This result is consistent with
previous findings (van der Marel and Cioni 2001, Nikolaev \etal 2004,
Persson \etal 2004, Pejcha and Stanek 2009, Koerwer 2009, Subramanian and
Subramaniam 2010, Haschke \etal 2012a, Subramanian and Subramaniam 2013,
van der Marel and Kallivayalil 2014, Deb and Singh 2014). We slice-up the
galaxy into distance intervals in Fig.~4 to see the details of this
tilt. Top three panels show LMC parts that are closer than 50~kpc, while
bottom three panels that are farther than 50~kpc (which is very close to
the mean distance to the LMC $d_{\rm LMC}=49.97$~kpc from Pietrzyñski \etal
2013). There is a clear difference between the top and the bottom row --
the closest LMC stars are located mainly in the eastern parts of the
galaxy, especially the eastern part of the bar and the northern arm, while
the farthest parts of the LMC are in the west. Moreover, the northern arm
seems to lie closer to us than the rest of the galaxy. The bar will be
discussed in detail in Section~5.
\begin{figure}[htb]
\includegraphics[width=12.5cm]{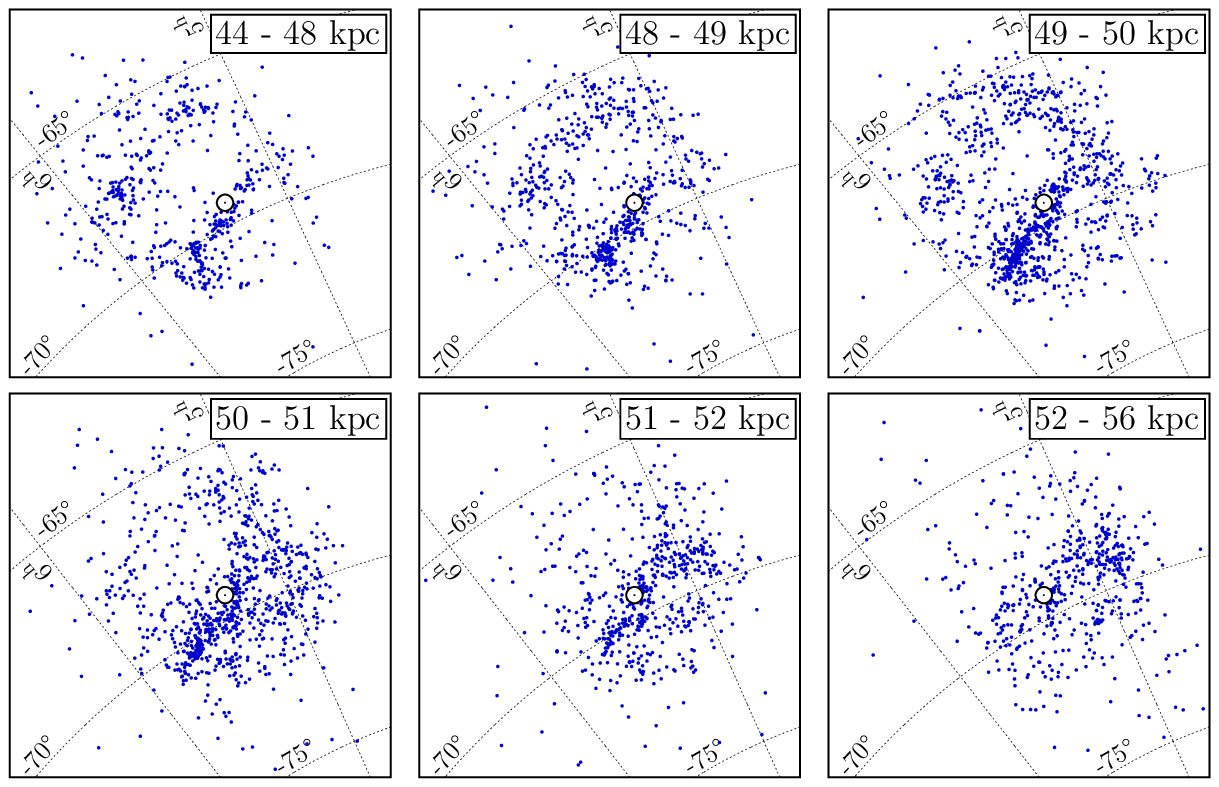}
\FigCap{Distance tomography of the LMC in the Hammer projection. Note that
  the {\it first} and the {\it last panel} distance range is 4~kpc, the
  {\it intermediate panels} -- 1~kpc. White circle marks the LMC center
  (van der Marel and Kallivayalil 2014).}
\end{figure}

\begin{figure}[p]
  \centerline{\includegraphics[width=10cm]{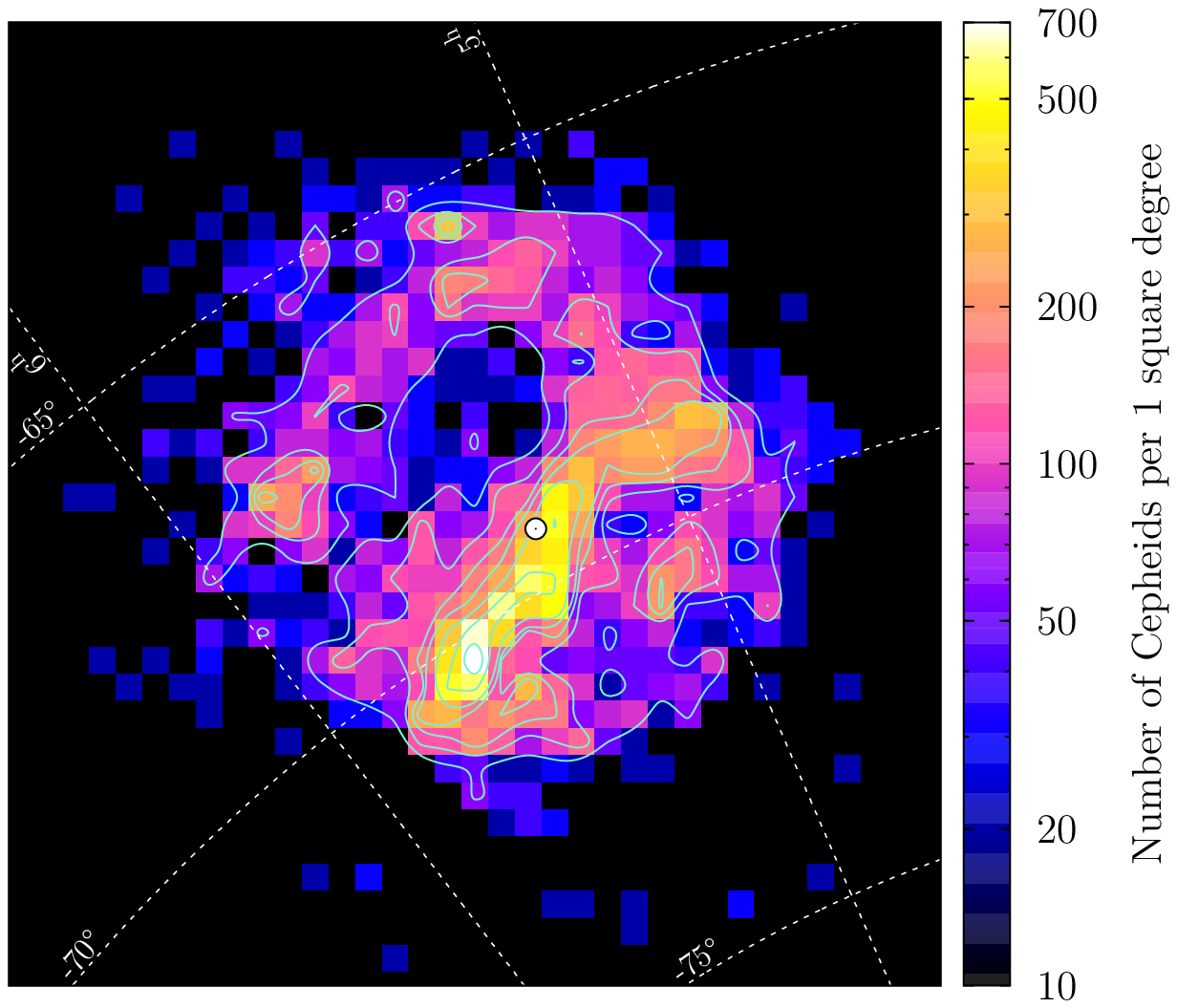}}
  \centerline{\includegraphics[width=10cm]{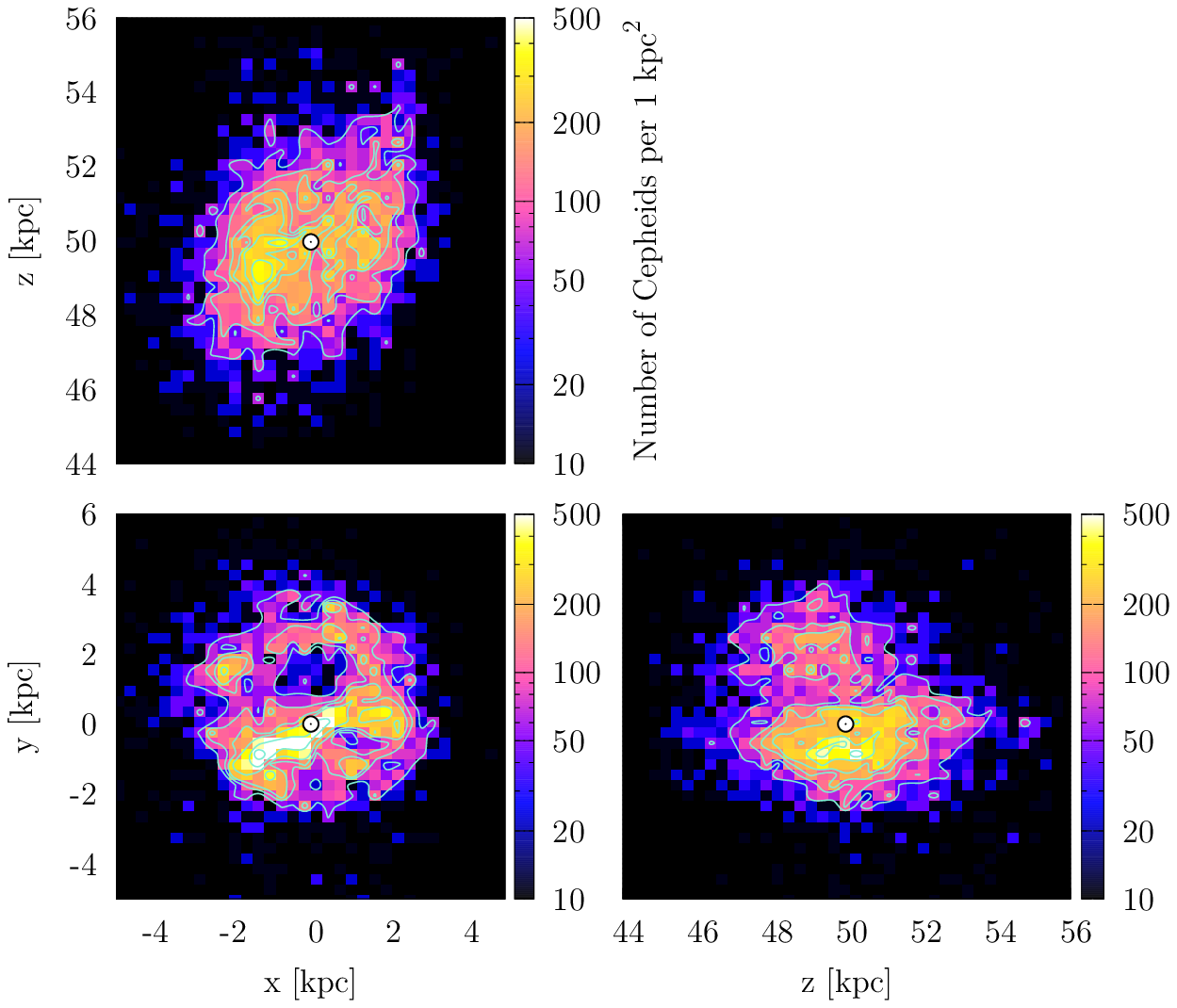}} 
  \FigCap{Cepheid density maps in the LMC with Cepheid column density
    contours.  {\it Top map:} Map in the Hammer projection. The bin size is
    0.0001 in units of Hammer projection coordinates $x_{\rm Hammer}$ and
    $y_{\rm Hammer}$ in both directions.  Contour levels are: 50, 100, 150,
    200, 300, 500, 700 Cepheids per 1 square degree.  {\it Bottom set of
    three maps:} Maps in the Cartesian coordinates projections with the
    {\it z} axis pointing toward the LMC center.  The bin size is 0.3~kpc
    in {\it x}, {\it y} and {\it z}. Contour levels on the {\it xy} plane
    are 50, 100, 150, 250, 500, 800, on the {\it xz} plane 50, 100, 150,
    200, 250, 300 and on the {\it yz} plane 50, 100, 150, 200, 300, 400
    Cepheids per 1~kpc$^2$.  The white circle marks the LMC center
    (Pietrzyñski \etal 2013, van der Marel and Kallivayalil 2014).}
\end{figure}
In Fig.~5 we show Cepheid column density maps. The top map is visualized in
the Hammer projection and the bottom three in the Cartesian planes {\it
  xy}, {\it xz} and {\it yz}, with the {\it z} axis pointing toward the LMC
center. The most prominent feature is the bar -- especially its eastern
part -- and the northern arm.  The northern arm shows a number of
overdensities: one is connected with the bar, another two are on the
northmost side of the LMC and the fourth one is at the tip of the arm.  We
also see two Cepheid overdensities in the southern part of the LMC, which
may indicate a presence of another arm. The larger of these overdensities
seems to be connected with and coming out of the bar at its east end --
this is also visible in the first panel of Fig.~4.  The other southern
overdensity is separated from the bar.

The bottom set of three maps in Fig.~5 shows bins in the Cartesian
projections, see figure caption for a full description.  The map showing
the {\it xy} plane is very similar to the top map. The bar has the largest
column density and its eastern part is the most prominent feature of the
galaxy. The northern arm and its overdensities, as well as the southern
structures, are also well distinguishable.  The {\it xz} plane (view ``from
the top'') shows that the inclination of the LMC is very evident. The
eastern part of the LMC lies closer to us and is more numerous than the
western part. The {\it yz} plane (view ``from the side'') shows two almost
separate parts: the northern and the southern, that comprise with the LMC
northern arm and the bar, respectively. This projection clearly shows that
the arm is closer to us than the LMC, as implied in Fig.~4.  On the other
hand, the southern part is at a similar distance as the mean LMC
distance. Contrary to previous studies (\eg Zhao and Evans 2000, Nikolaev
\etal 2004, Subramanian and Subramaniam 2013, van der Marel and
Kallivayalil 2014 and numerical model of the off-center bar in Bekki 2009
and Besla \etal 2012), we do not see that the bar is located closer to us
than the LMC.

\begin{figure}[h]
\centerline{\includegraphics[width=12.3cm]{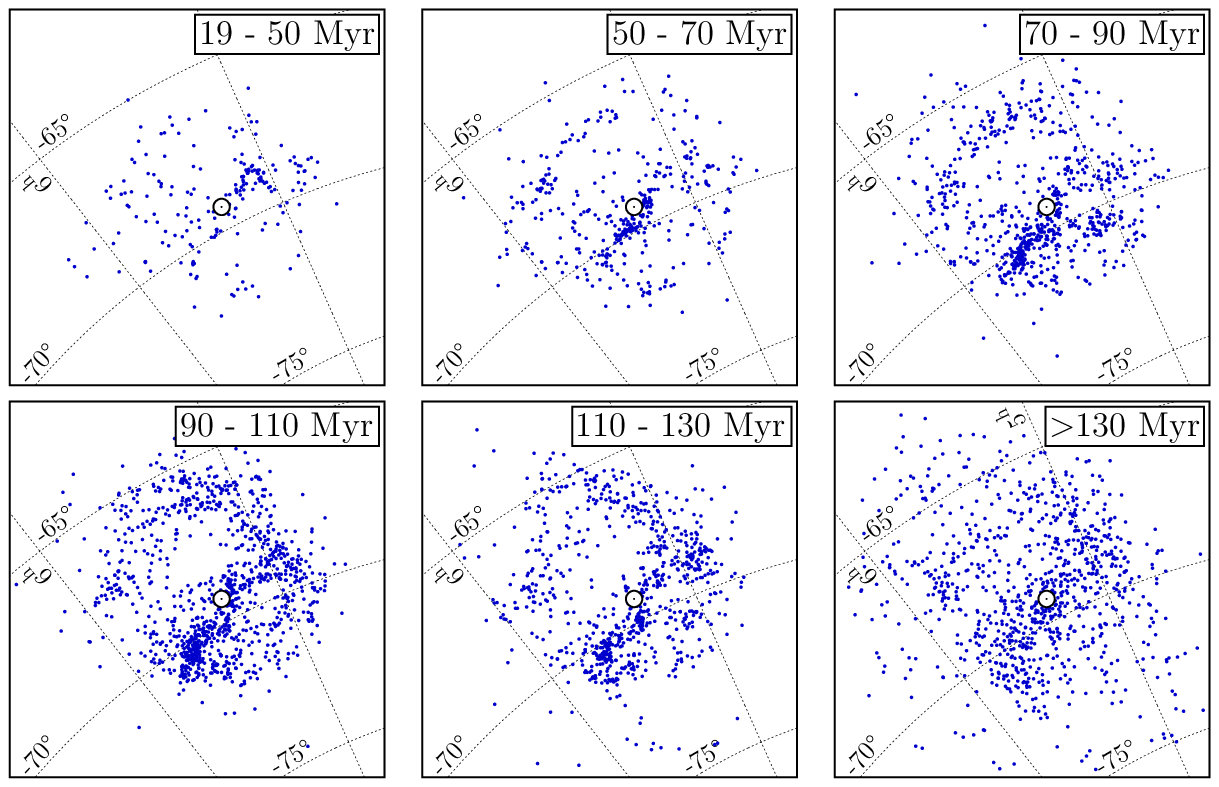}}
\FigCap{Age tomography of the LMC using the relation from Bono \etal (2005)
  for a constant metallicity $Z=0.01$. The maps are in the Hammer
  projection.  Note that the {\it first panel} shows an interval of 51~Myr,
  {\it last} -- 263~Myr, {\it the other ones} -- 20~Myr.  White circle
  marks the LMC center (van der Marel and Kallivayalil, 2014).}
\end{figure}
\vspace*{-7pt}
\subsection{Ages}
We estimated ages of the LMC Cepheids using the period--age relation from
Bono \etal (2005) for a constant metallicity $Z=0.01$. Some studies
suggest that LMC has a metallicity gradient (Cioni 2009, Feast \etal
2010, Wagner-Kaiser and Sarajedini 2013), but a recent study by Deb and
Singh (2014) shows that there is no such gradient or it is too small to be
detected with techniques used.

The on-sky distribution of Cepheids in different age intervals is presented
in Fig.~6.  Most of the stars fall into the age range of 50--130~Myr. The
youngest Cepheids are found in the western part of the bar at
$\alpha\approx5\uph$ and are younger than 50~Myr. In the age interval of
50--70~Myr the central part of the bar emerges. Then the eastern part of
the bar shows up along with the western part and the northern arm.  The
eastern and western areas of the bar were formed at similar times and thus
should be treated as parts of one coherent structure. Cepheids older than
130~Myr are scattered along the bar and the arm and are spread all over the
LMC disk.

Soszyñski \etal (2015) noticed that there is a difference between the
distributions of fundamental and first-overtone Cepheids in the LMC, such
that 1O stars are more spread than F-mode stars (see their Fig.~4). This
can be explained by age differences between these types -- the 1O Cepheids
are slightly older and so had time to spread.

\subsection{Substructures}
To investigate properties of the bar, the arm, and other structures of the
LMC we divided the galaxy into several regions shown in Fig.~7.  The left
panel illustrates selection areas for main structures: the whole bar and
the whole arm as well as two southern regions. We further divided the bar
and the arm each into two subregions -- eastern and western bar, and
northern arm 1 and northern arm 2, as shown in the right panel. Basic
parameters of all substructures, such as the median distance and age,
standard deviations and number of stars in each group, are listed in
Table~4.
\renewcommand{\TableFont}{\footnotesize}
\MakeTableee{l|cc|cc|r}{12.5cm}{Characteristics of the LMC substructures}
{\hline
\douprule
Substructure & $\langle{\rm dist}\rangle$ [kpc] & $\sigma_{\rm dist}$ [kpc]& $\langle{\rm age}\rangle$ [Myr] & $\sigma_{\rm age}$ [Myr]& \multicolumn{1}{|c}{N} \\
\hline
\uprule
All Cepheids      & 49.93 &  1.79 & 104 & 53 & 4222 \\
Bar               & 50.03 &  1.74 & 100 & 48 & 1662 \\
Eastern Bar       & 49.86 &  1.65 & 100 & 49 & 1318 \\
Western Bar       & 51.03 &  1.82 & 104 & 45 & 344 \\
Northern Arm      & 49.39 &  1.66 & 106 & 48 & 965 \\
Northern Arm 1    & 49.43 &  1.70 & 105 & 50 & 820 \\
Northern Arm 2    & 49.13 &  1.35 & 108 & 34 & 143 \\
Southern Region 1 & 49.96 &  1.73 & 106 & 46 & 236 \\
\dorule
Southern Region 2 & 50.78 &  1.39 & 101 & 52 & 190 \\
\hline
\noalign{\vskip2pt}
\multicolumn{6}{p{10cm}}{The table lists median distance and age together with standard deviations, and a number of stars in each substructure.}
}
\begin{figure}[htb]
\includegraphics[width=12.5cm]{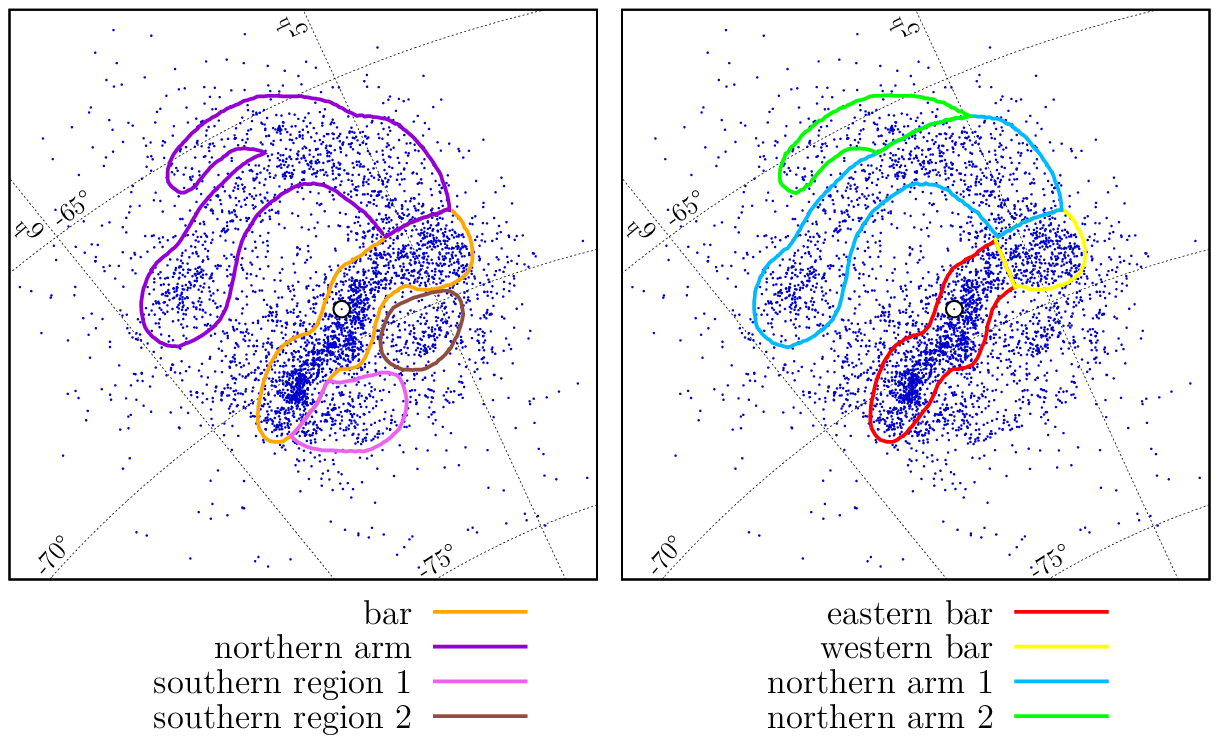}
\FigCap{The maps show the LMC Cepheids in the Hammer projection. {\it
    Left:} Main regions are presented: the whole bar, the whole northern
  arm and two southern regions. {\it Right:} The map shows divisions of the
  bar and the arm into two subregions: eastern and western bar, northern
  arm 1 and northern arm 2. White point with dot in the middle marks the
  LMC center (van der Marel and Kallivayalil 2014).}
\end{figure}

When constructing the selection areas for each structure we followed the
density contours for binned data shown in Fig.~5. The choice was also based
on distributions of stars in different age intervals (see Fig.~6). The
age-space distributions were discussed in detail in Section~4.2.
Here we concentrate on justification of the selected regions and their
properties.

The selection of the bar area was performed in a few stages. The density
contours suggest that the bar may consist of two parts: eastern, making up
almost the whole bar in terms of star counts, and western. The eastern bar
which is regarded as the ``classical'' LMC bar (see Fig.~14 in Nikolaev
\etal 2004 and Figs.~1 and 7 in Haschke \etal 2012a) is the densest and the
brightest part of the LMC. It is also located about 0.5~kpc closer than the
rest of this galaxy (\eg Zhao and Evans 2000, Cioni \etal 2000, Nikolaev
\etal 2004, Subramanian and Subramaniam 2013, van der Marel and
Kallivayalil 2014 and numerical models of the off-center bar in Bekki 2009
and Besla \etal 2012). However, Fig.~4 suggests that the entire bar should
consist of both the eastern and the western part. There is a fairly
continuous band of stars between the parts and there is no significant
break between these parts at any of the distance slices. Even though the
first two panels of Fig.~4 show mainly the eastern bar, the third map
(distance interval 49--50~kpc) shows a western counterpart. At larger
distances we see that the eastern area fades and the western is more
visible. The age-tomography (Fig.~6) leads to very similar conclusions: the
maps showing age intervals 90--110~Myr and 110--130~Myr represent the most
evident connection between the eastern and western area of the
bar. Moreover, the dynamical center of the LMC, marked in Fig.~7 with a
white circle, is located in the middle of the whole bar, not its eastern
part.

\begin{figure}[htb]
\centerline{\includegraphics[width=11cm]{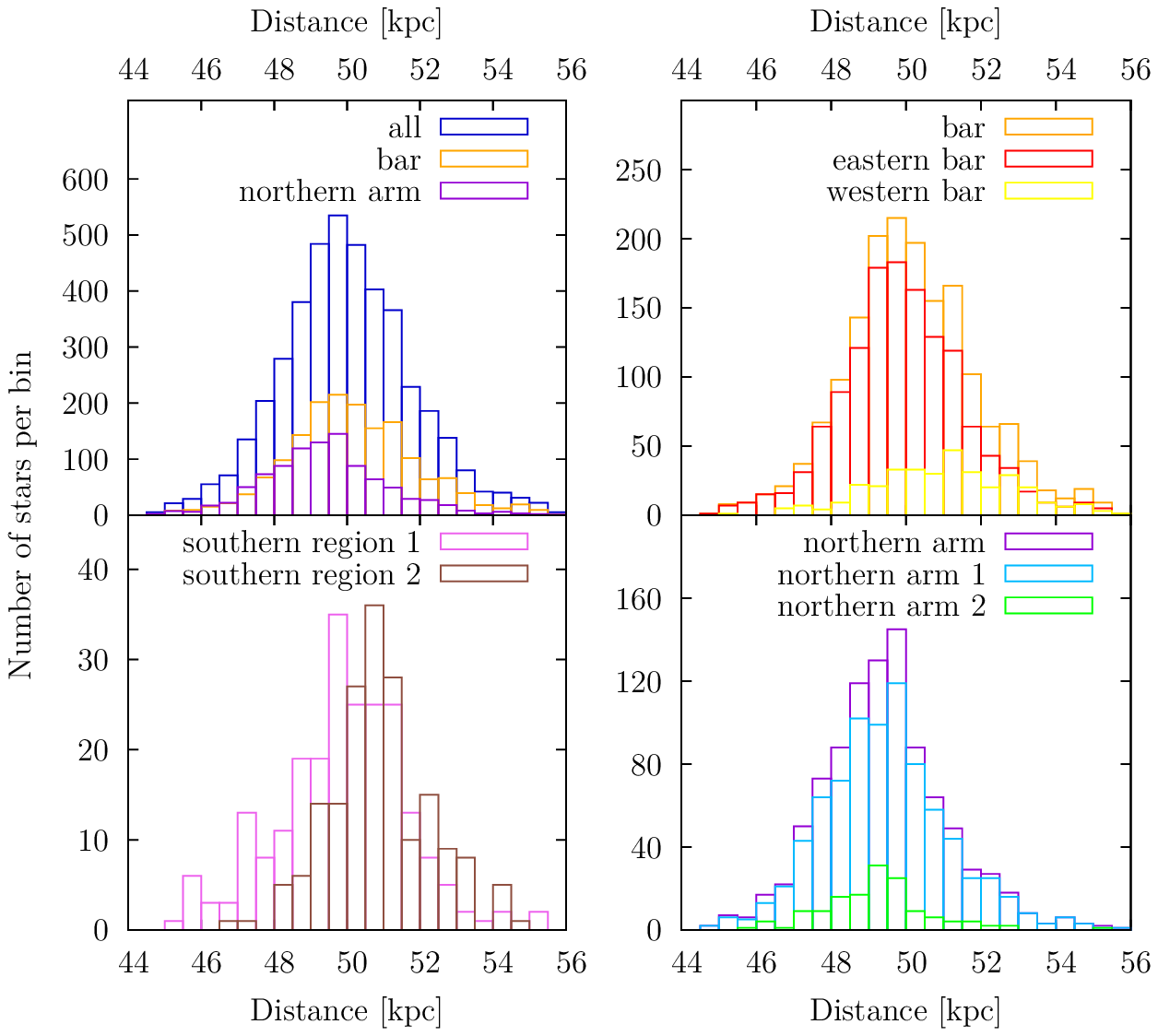}}
\FigCap{Distance histograms of the selected regions in the LMC. {\it Left
  panel, top:} Comparison of all LMC Cepheids with the main structures --
  the bar and the northern arm. {\it Left panel, bottom:} Southern regions
  1 and 2. {\it Right panel, top:} Comparison of the whole bar with its
  eastern and western parts. {\it Right panel, bottom:} Comparison of the
  whole northern arm with its parts 1 and 2.}
\end{figure}
A histogram showing the comparison of the distance distribution in the
whole LMC and the bar (as well as the northern arm) is in the top left
panel of Fig.~8. We perform a series of Kolmogorov-Smirnov (KS) tests for
the null hypothesis that the two samples come from the same distribution,
and the test results for various samples are listed in Table~5. In the case
of the whole LMC and the bar we obtained $D=0.039$ and a $\pvalue=0.048$.
This means that the hypothesis can be rejected at a significance level
$\alpha=0.05$. However, according to Sellke \etal (2001), the error rate
associated with a $\pvalue$ of $\approx0.05$ is at least 23\% and typically
$\approx50\%$ (which is the probability that a true null hypothesis has
been rejected). In the case of the $\pvalue=0.01$, the error rate is at
least 7\% and typically $\approx15\%$, thus in the following analysis we
will assume that the null hypothesis can be rejected only if $\pvalue\leq
0.01$. According to the KS test results, and median distances from Table~4
we again see that the bar does not lie closer to us than the LMC, when
defined as described in the previous paragraph. The top right panel of
Fig.~8 shows a histogram of the entire bar and separately its eastern and
western parts. Here we can see that the eastern part does lie closer to us
than the the western part, which is supported by their median distances
(49.86~kpc and 51.03~kpc, respectively) and the KS test results at
significance level $\alpha=0.001$ ($D=0.287$, $\pvalue=0$). If we treat the
bar in a ``classical'' way, \ie as its eastern part, then there is no
strong evidence that it is located closer to us than the LMC (the offset is
only 0.07~kpc, see Table~4). Also, the significance level at which we could
reject the hypothesis of the two distributions coming from the same sample
is only $\alpha=0.1$ ($D=0.039$, $\pvalue=0.089$).

\begin{figure}[htb]
\centerline{\includegraphics[width=11cm]{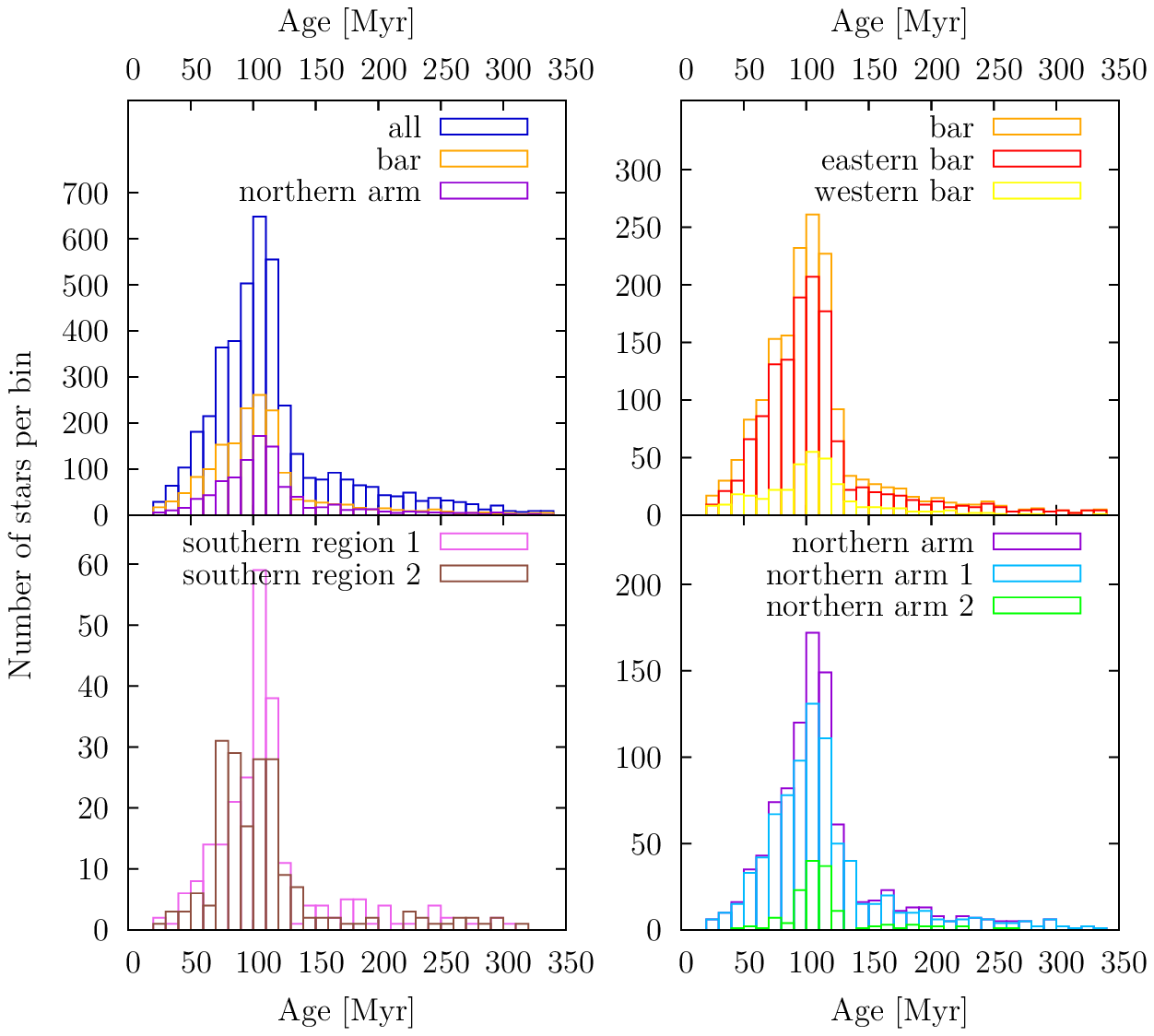}}
\FigCap{Age histograms of the selected regions in the LMC. {\it Left panel,
  top:} Comparison of all LMC Cepheids with the main structures -- the
  bar and the northern arm. {\it Left panel, bottom:} Southern regions 1
  and 2. {\it Right panel, top:} Comparison of the whole bar with its
  eastern and western parts. {\it Right panel, bottom:} Comparison of the
  whole northern arm with its parts 1 and 2.}
\end{figure}
The age histograms in the top left panel of Fig.~9 show that Cepheids' age
distribution in the bar is fairly similar to the age distribution of the
entire galaxy, as supported by median ages in Table~4, but the KS test
results presented in Table~5 allow us to reject this hypothesis at a
significance level $\alpha=0.001$ ($D=0.069$, $\pvalue=0$). The top
right panel suggests that the western part of the bar is slightly older
than the eastern part, but since $\pvalue=0.042$, we cannot reject
that they come from the same age distribution. This further supports our
choice of the bar region.

\renewcommand{\TableFont}{\footnotesize}
\MakeTable{|ll|ccc|ccc|}{12.5cm}{Kolmogorov-Smirnov test results in the LMC}
{\hline
 &  & \multicolumn{3}{c|}{DISTANCE} & \multicolumn{3}{c|}{AGE}\uprule \\
Sample 1 & Sample 2 & $D$ & $\pvalue$ & $\alpha^*$ & $D$ & $\pvalue$ & $\alpha^*$\dorule \\
\hline
\uprule
all     &  bar    &  0.039 & 0.048 & 0.050 & 0.069 & 0.000 & 0.001 \\
all     &  bar-E  &  0.039 & 0.089 & 0.100 & 0.079 & 0.000 & 0.001 \\
all     &  arm-N  &  0.165 & 0.000 & 0.001 & 0.041 & 0.133 & ----- \\
arm-N   &  bar    &  0.193 & 0.000 & 0.001 & 0.098 & 0.000 & 0.001 \dorule\\
\hline
\uprule
bar     &  bar-E  &  0.060 & 0.009 & 0.025 & 0.017 & 0.980 & ----- \\
bar     &  bar-W  &  0.227 & 0.000 & 0.001 & 0.066 & 0.159 & ----- \\
bar-E   &  bar-W  &  0.287 & 0.000 & 0.001 & 0.083 & 0.042 & 0.050 \dorule\\
\hline
\uprule
SR1     &  SR2    &  0.295 & 0.000 & 0.001 & 0.139 & 0.031 & 0.050 \dorule\\
\hline
\uprule
arm-N   &  arm-N1 &  0.025 & 0.942 & ----- & 0.031 & 0.781 & ----- \\
arm-N   &  arm-N2 &  0.146 & 0.009 & 0.025 & 0.173 & 0.001 & 0.005 \\
arm-N1  &  arm-N2 &  0.171 & 0.001 & 0.005 & 0.204 & 0.000 & 0.001 \dorule\\
\hline
\noalign{\vskip3pt}
\multicolumn{8}{p{10cm}}{$^*\alpha$ is a significance level at which a null
hypothesis that the two samples come from the same distribution can be
rejected. No value means that $\alpha\ge0.100$ and the hypothesis cannot
be rejected. Due to our strict approach we treat values only below
$\alpha=0.010$ as significant and allowing us to reject the hypothesis.}
}

The northern arm selection area was based on density contours (Fig.~5). We
divided the arm into two parts that we named northern arm 1 and northern
arm 2 (hereafter NA1 and NA2). The NA1 is the most prominent part of the
whole northern arm. It is connected with the western part of the bar and
stretches out to the northern and eastern side of the LMC. The NA2 is
located in the northmost part of the LMC and is connected with NA1. It is
visible as the brightest overdensity in the northern part of
Fig.~5. Soszyñski \etal (2015) noticed that it is only visible in
fundamental mode Cepheids.

The distance histogram in the top left panel of Fig.~8 shows that the
northern arm is located closer to us than the whole LMC at a significance
level $\alpha=0.001$ (see Table~5). The bottom right panel compares
distance distributions of NA1 and NA2. Their distances are consistent with
an overall distance of the northern arm, but the KS test shows a difference
in their distributions at level $\alpha=0.005$. On the other hand, the age
histograms and KS test results in Table~5 lead to a conclusion that the arm
is slightly older than the bar (top left panel of Fig.~9), but there is no
age difference between the northern arm and the LMC.

The first map in the top panel of Fig.~4 suggests that there might be another
arm in the southern part of the LMC. It seems to be connected with the bar at
its south-east end.

We subdivide this region into two parts: southern region 1 (SR1) and
southern region 2 (SR2) shown in Fig.~7. Their mean distances (Table~4) are
consistent with the inclination of the LMC disk. The SR1, which is located
in the eastern part of the LMC, is also closer to us than SR2, that is
located in the western part of the galaxy, at significance level
$\alpha=0.001$ (Table~5). Interestingly, SR2 seems to be younger than SR1,
but the significance of this claim is low ($\alpha=0.05$), thus we do not
treat this result as relevant.

\subsection{Plane Fitting}
We performed a three-dimensional plane fitting to the LMC Cepheids as
described in Section~3.4. We used Cartesian coordinates {\it x}, {\it y},
{\it z} although in the plane-fitting model the coordinate system center is
placed in the LMC center and {\it z} axis points in the opposite direction
than on our map projections. We separately fit CCs in the bar, in the
northern arm and for the entire LMC. The three-dimensional selection areas
for the bar and the arm are shown in Fig.~10. We do realize that fitting a
simple plane is a great oversimplification, especially in the case of the
bar, but the scope of this paper is a rough estimation of the global
parameters for which a simple plane fitting is sufficient.
\begin{figure}[p]
  \centerline{\includegraphics[width=11cm]{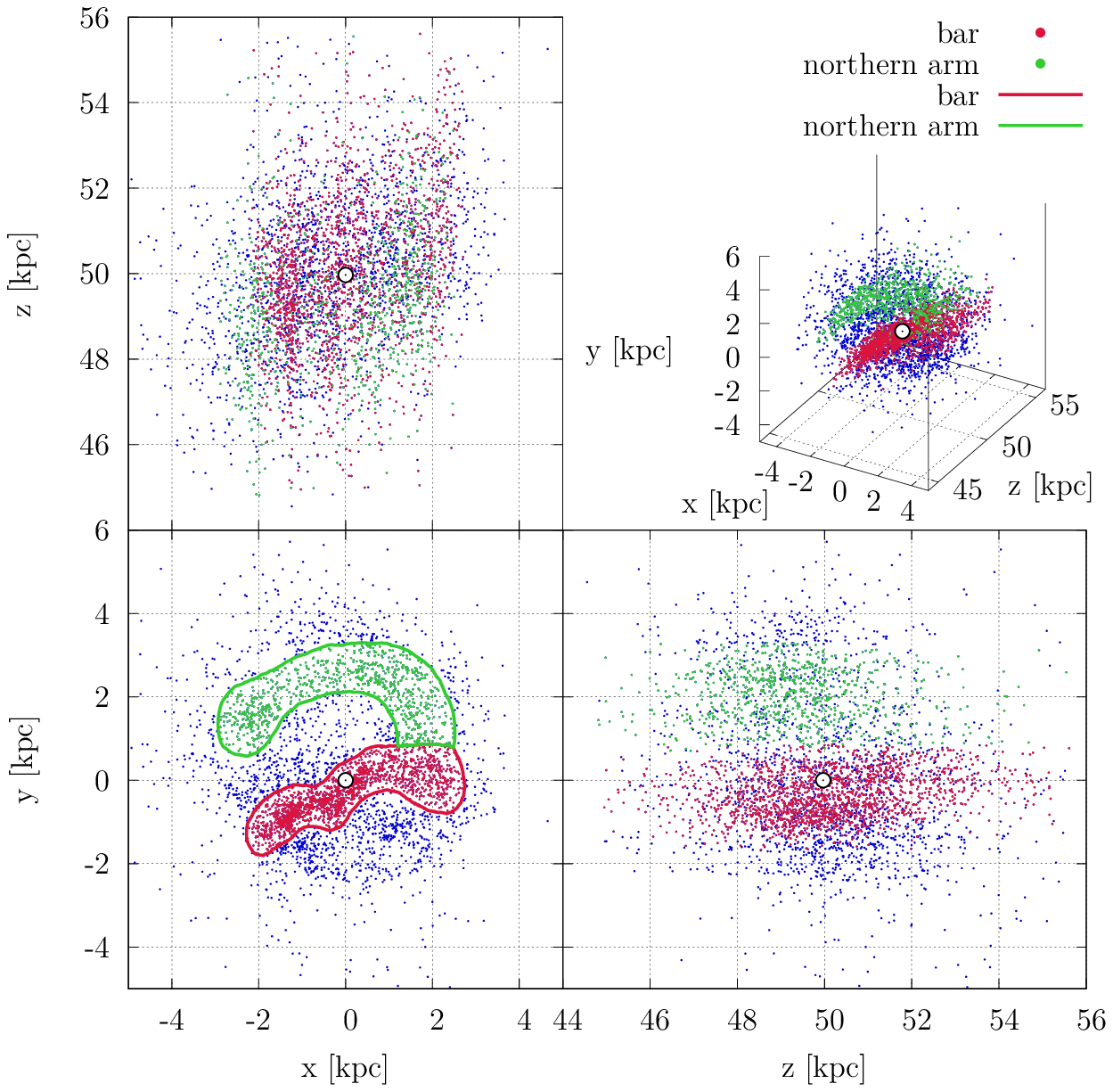}}
  \FigCap{Three-dimensional map of the CCs in the LMC in Cartesian
    coordinates with the {\it z} axis pointing toward the LMC center. Blue
    dots represent the entire LMC Cepheid population, red dots the bar and
    green dots northern arm. The {\it xy} plane shows the selection regions
    for the bar and for the northern arm for plane-fitting. White circle
    marks the LMC center (Pietrzyñski \etal 2013, van der Marel and
    Kallivayalil 2014).}  
\vspace*{3mm}
  \centerline{\includegraphics[width=7cm]{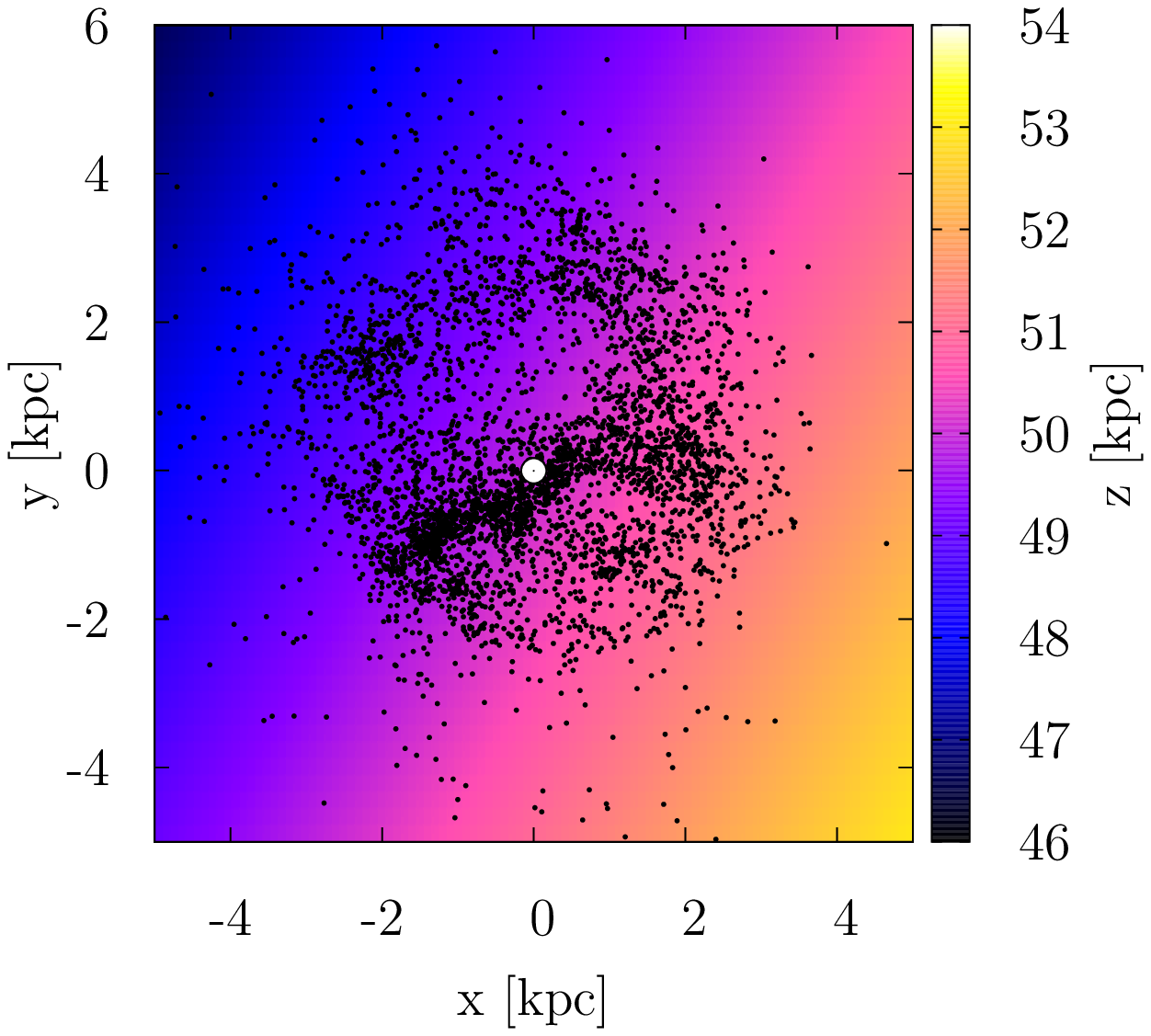}}
  \FigCap{Distance-gradient of the best-fit plane for the entire LMC
    (color-coded) in Cartesian {\it xy} coordinates with the {\it z} axis
    pointing toward the LMC center. Black dots show all LMC Cepheids. White
    circle marks the LMC center (van der Marel and Kallivayalil 2014).}
\end{figure}

The best-fit parameters are listed in Table~6, where $a$, $b$ and $c$ are
plane equation coefficients, $i$ and P.A. are inclination and position
angle respectively. There are separate sets of parameters for all LMC
Cepheids, for all except the bar, for the bar, and for the northern
arm. All fits have {\it rms} values of about 1.5~kpc, which is a result of
the inaccuracy of distance determination.

\MakeTable{|l|cr@{}lcc|}{12.5cm}{Best-fit parameters of the three-dimensional plane fitting procedure}
{\hline
\douprule
LMC data       & $a$              & \multicolumn{2}{c}{$b$} & $c$ [kpc]  & $N$ \\
 \hline
All Cepheids   & $-0.395\pm0.014$ & 0   &$.215\pm0.013$     & $-0.005\pm0.021$ & 4190 \\
All except bar & $-0.354\pm0.016$ & 0   &$.237\pm0.014$     & $-0.013\pm0.031$ & 2458 \\
Bar            & $-0.414\pm0.039$ & $-0$&$.048\pm0.095$     & $-0.094\pm0.045$ & 1731 \\
Northern arm   & $-0.378\pm0.032$ & 0   &$.571\pm0.082$     & $-0.463\pm0.170$ & 756 \\
\hline 
\douprule
LMC data       & {\it i} & \multicolumn{2}{c}{P.A.} & $\chi^2$/dof & {\it rms} [kpc] \\ \hline
All Cepheids   & $24\zdot\arcd2\pm0\zdot\arcd7$ & 1&5$1\zdot\arcd4\pm1 \zdot\arcd7$ & 1.355 & 1.5 \\
All except bar & $23\zdot\arcd1\pm0\zdot\arcd8$ & 1&4$6\zdot\arcd1\pm2 \zdot\arcd0$ & 1.323 & 1.5 \\
Bar            & $23\zdot\arcd1\pm1\zdot\arcd5$ & 1&8$7\zdot\arcd2\pm12\zdot\arcd6$ & 1.376 & 1.5 \\
Northern arm   & $34\zdot\arcd4\pm2\zdot\arcd9$ & 1&2$3\zdot\arcd8\pm3 \zdot\arcd8$ & 1.163 & 1.2 \\
\hline
\noalign{\vskip3pt}
\multicolumn{6}{c}{The coefficients were calculated using the Markov chain Monte Carlo method.}
}

In the case of all LMC Cepheids, we obtain $i=24\zdot\arcd2\pm
0\zdot\arcd7$ and ${\rm P.A.}=151\zdot\arcd4\pm1\zdot\arcd7$ that correlate
well with values from the literature (see comparison in Table~7). The
parameter $c$, which is an offset of the fitted plane from the LMC center
along {\it z} axis in kpc, is very small and consistent with the two
centers being identical. Fig.~11 shows the {\it z} coordinate gradient and
therefore the direction of LMC's tilt.

The fit to all Cepheids except those in the bar gives identical values of
$i$ and P.A. (within $1\sigma$ errors for $i$ and $2.26\sigma$ for P.A.),
showing that the bar does not influence the fit. This is also consistent
with the result from Subramanian and Subramaniam (2013) who analyzed the
red clump stars in the LMC and found that the bar is a co-planar structure,
although it may be offset from the plane by up to 0.5~kpc in the direction
of the observer. This offset is not reflected in parameter $c$ of our fit,
which for the bar is $-0.094\pm0.045$~kpc and this value is statistically
insignificant within $3\sigma$ uncertainty. As discussed in previous
sections, this is an effect of the bar selection criteria.

\MakeTableee{l@{\hspace{7pt}}c@{\hspace{7pt}}c@{\hspace{5pt}}l}{12.5cm}{LMC disk parameters from the literature}
{\hline
\noalign{\vskip3pt}
\multicolumn{4}{l}{Cepheids and young population} \\ 
\noalign{\vskip3pt}
\hline
\noalign{\vskip3pt}
Reference                             & {\it i} & P.A. & Data \\ 
\noalign{\vskip3pt}
\hline
\noalign{\vskip3pt}
This work: all                        & $24\zdot\arcd2{\pm}0\zdot\arcd7$ & $151\zdot\arcd4{\pm}1\zdot\arcd7$  & OGLE-IV CCs \\
This work: bar only                   & $23\zdot\arcd1{\pm}1\zdot\arcd5$ & $187\zdot\arcd2{\pm}12\zdot\arcd6$ & OGLE-IV CCs \\
This work: arm only                   & $34\zdot\arcd4{\pm}2\zdot\arcd9$ & $123\zdot\arcd8{\pm}3\zdot\arcd8$  & OGLE-IV CCs \\ 
\noalign{\vskip3pt}
\arrayrulecolor{lightgray}\hline
\noalign{\vskip3pt}
Caldwell and Coulson (1986)           & $29\arcd{\pm}7\arcd$             & $142\arcd{\pm}8\arcd~$~            & Cepheids \\
Laney and Stobie (1986)               & $45\arcd{\pm}7\arcd$             & $145\arcd{\pm}17\arcd$             & Cepheids \\
van der Marel and Cioni (2001)        & $34\zdot\arcd7{\pm}6\zdot\arcd2$ & $122\zdot\arcd5{\pm}8\zdot\arcd3$  & AGB stars \\
Nikolaev \etal (2004)                 & $30\zdot\arcd7{\pm}1\zdot\arcd1$ & $151\zdot\arcd0{\pm}2\zdot\arcd4$  & Cepheids \\
Persson \etal (2004)                  & $27\zdot\arcd0{\pm}6\zdot\arcd0$ & $127\arcd{\pm}10\arcd$             & Cepheids \\
Haschke \etal (2012a)                 & $32\arcd{\pm}4\arcd$             & $115\arcd{\pm}15\arcd$             & OGLE-III CCs \\
van der Marel and Kallivayalil (2014) & $26\zdot\arcd2{\pm}5\zdot\arcd9$ & $154\zdot\arcd5{\pm}2\zdot\arcd1$  & PM + young pop. LOS velocity \\ 
\arrayrulecolor{black}\noalign{\vskip3pt}
\hline
\noalign{\vskip3pt}
\multicolumn{4}{l}{Other tracers} \\ 
\noalign{\vskip3pt}
\hline
\noalign{\vskip3pt}
Reference                             & {\it i} & P.A. & Data \\ \hline
Koerwer (2009)                        & $23\zdot\arcd5{\pm}0\zdot\arcd4$ & $154\zdot\arcd6{\pm}1\zdot\arcd2$  & Red clump \\
Subramanian and Subramaniam (2010)    & $23\zdot\arcd0{\pm}0\zdot\arcd8$ & $163\zdot\arcd6{\pm}1\zdot\arcd5$  & OGLE-III RR~Lyr \\
Subramanian and Subramaniam (2010)    & $37\zdot\arcd4{\pm}2\zdot\arcd3$ & $141\zdot\arcd2{\pm}3\zdot\arcd7$  & MCPS data \\
Rubele \etal (2012)                   & $26\zdot\arcd2{\pm}2\zdot\arcd0$ & $129\zdot\arcd1{\pm}13\zdot\arcd0$ & VMC data \\
Haschke \etal (2012a)                 & $32\arcd{\pm}4\arcd$             & $116\arcd{\pm}18\arcd$             & OGLE-III RR~Lyr \\
Subramanian and Subramaniam (2013)    & $25\zdot\arcd7{\pm}1\zdot\arcd6$ & $141\zdot\arcd5{\pm}4\zdot\arcd5$  & Red clump outer disk ($r>3\arcd$) \\
van der Marel and Kallivayalil (2014) & $39\zdot\arcd6{\pm}4\zdot\arcd5$ & $147\zdot\arcd4{\pm}10\zdot\arcd0$ & Proper motion (PM) data \\
van der Marel and Kallivayalil (2014) & $34\zdot\arcd0{\pm}7\zdot\arcd0$ & $139\zdot\arcd1{\pm}4\zdot\arcd1$  & PM + old pop. LOS velocity \\
Deb and Singh (2014)                  & $24\zdot\arcd20$                 & $176\zdot\arcd01$                  & OGLE-III RR~Lyr (ellipsoid) \\
Deb and Singh (2014)                  & $36\zdot\arcd43$                 & $149\zdot\arcd08$                  & OGLE-III RR~Lyr (plane) \\
\noalign{\vskip3pt}
\hline
}

The fit to the northern arm Cepheids reveals a different nature of this
distribution. Both the inclination and position angle are inconsistent with
the literature within $3\sigma$ errors. The angle between the best-fit
planes for the LMC disk and the northern arm is about $40\arcd$. The $c$
parameter indicates that the northern arm is shifted by up to $-0.463\pm
0.170$~kpc (significant within $3\sigma$ errors) with respect to the LMC
center and thus it is located closer to us. This is consistent with
conclusions from previous sections.

Table~7 presents a comparison of our results with values from the
literature. The inclination and position angle for the whole LMC sample are
consistent with most of the results for young stars within the errors,
although $i$ is the lowest of all from Cepheid and young population
studies. On the contrary, the P.A. is well correlated with higher values.
Surprisingly, there is a significant difference between our results based
the on the OGLE-IV data, and results of Haschke \etal (2012a) who used the
OGLE-III Cepheids. As was already mentioned, the OGLE-III collection of CCs
did not contain most of the the northern arm and the southern structures.
This would indicate that the fit to the OGLE-III data should yield similar
results as our bar-only sample. The case is totally opposite -- our
inclination angle for the bar only is much lower than that of Haschke \etal
(2012a), while the P.A. is much higher. To check their {\it i} and
P.A. values we selected a similar sample from our data. We picked the
F-mode Cepheids located in OGLE-IV fields coinciding with OGLE-III
fields. Our fitting procedure resulted in values very similar to those
obtained for the entire LMC OGLE-IV Cepheid sample.

Results presented in this paper are also consistent with the parameter
values for the intermediate-age and older stellar populations (the second
part of Table~7).

\Section{The Small Magellanic Cloud}
\subsection{Three-Dimensional Structure}
The three-dimensional structure of the SMC is shown in Fig.~2. The galaxy
is elongated almost along the line of sight and its longitudinal dimension
(along the {\it z} axis) is about 4--5 times greater than transverse
dimensions in both {\it x} and {\it y} coordinates. This is perfectly
consistent with the latest findings by Scowcroft \etal (2016). The SMC
shape is best described as an extended ellipsoid with additional off-axis
structures that are also ellipsoidal. Note that the Wing is not clearly
visible in our data although in Figs.~2 and~3 we do see some Cepheids
located in that area. On the other hand, CCs are distributed all around the
SMC.

To show the change of shape of the SMC with increasing distance we have
performed the distance tomography. Fig.~12 shows sections of this galaxy in
different distance intervals. The closest part of the SMC ($d<59$~kpc) has
a round shape on the sky. The farther we look the less symmetrical it
becomes. Moreover, the Cepheids seem to move away from the dynamical center
of the SMC, marked with a white circle, to the south-western direction.

The second and the third map in the top row reveal an additional
substructure located in the north, that fades at a distance of about
65~kpc. At a similar distance range another substructure appears in the
south-west and is best visible on the second and the third map in the
bottom row.
\begin{figure}[htb]
\centerline{\includegraphics[width=12cm]{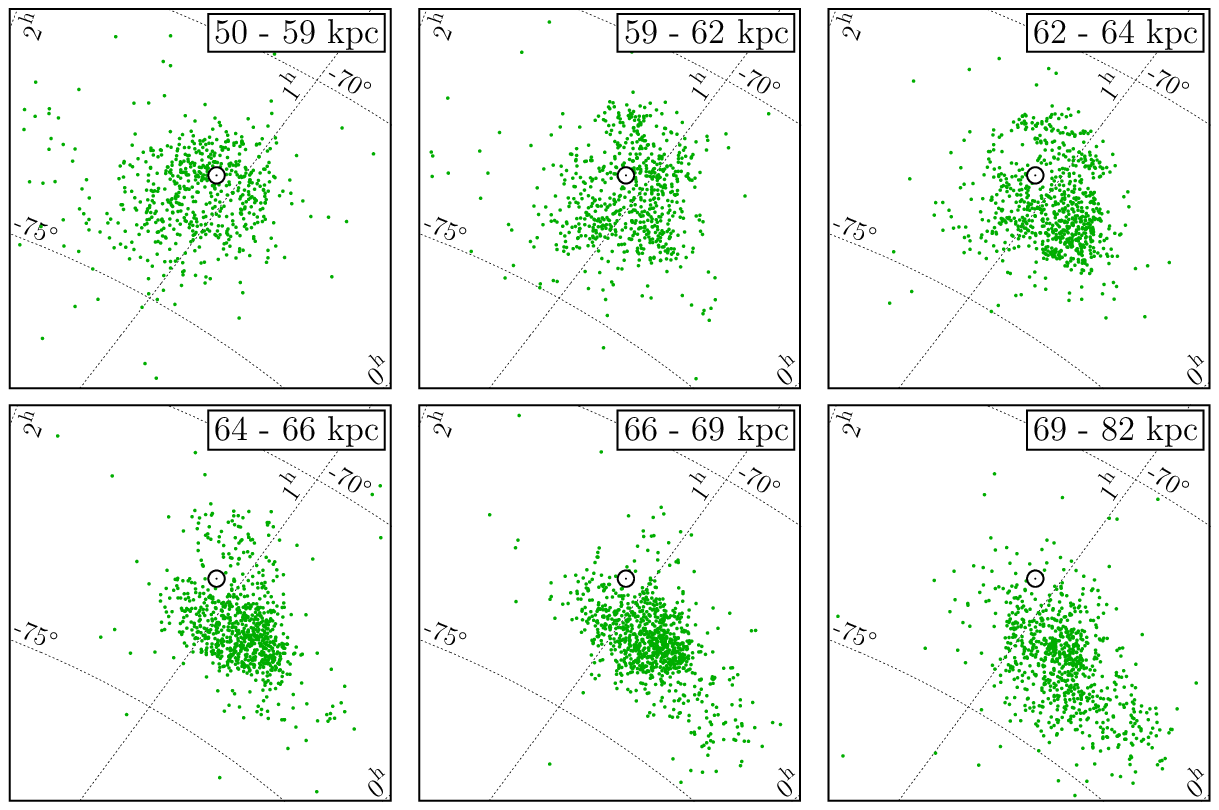}}
\FigCap{Distance tomography of the SMC in the Hammer projection. Note that
  the distance intervals are, starting from the upper left, 9, 3, 2, 2, 3,
  13~kpc. White circle marks the SMC center (Stanimiroviæ \etal 2004).}
\end{figure}

To better visualize the SMC subtle structures we binned the data both in
the Hammer projection and in the Cartesian space projections. The top map
in Fig.~13 shows the on-sky projection of the binned data with stellar
density contours overplotted. Interestingly, the higher density contours
omit the dynamical SMC center. We can see that the SMC is actually
heart-shaped with a curved tail in its south-western part. The top of the
``heart'' also suggests the existence of an additional substructure. This
part and the tail in the south-west were not clearly visible it the
OGLE-III Cepheid data (compare with Fig.~1 from Haschke \etal 2012b).

\begin{figure}[p]
\centerline{\includegraphics[width=10cm]{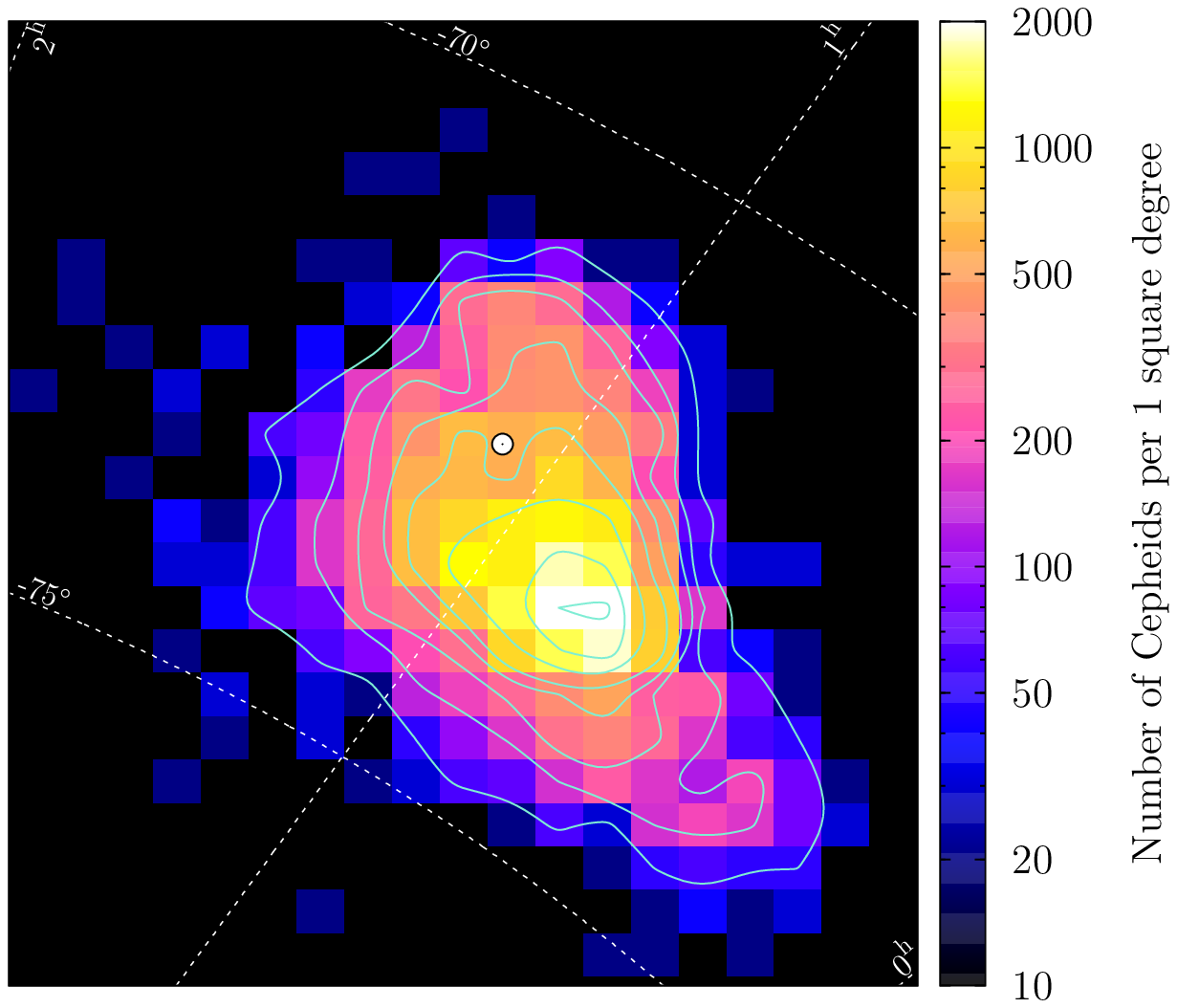}}
\centerline{\includegraphics[width=10cm]{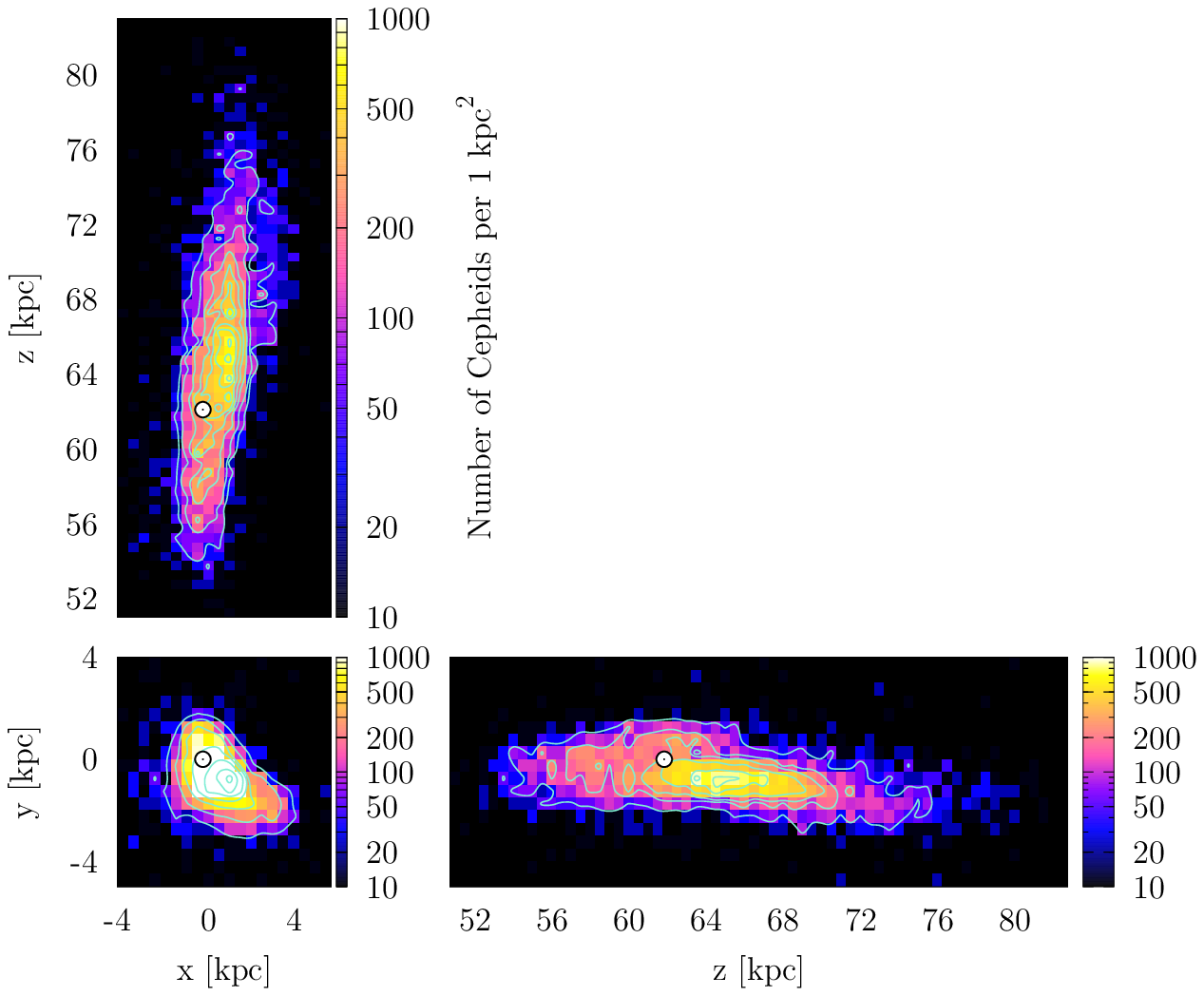}} 
\FigCap{Cepheid density in the SMC with Cepheid column density
  contours. {\it Top map:} Map in the Hammer projection. The bin size is
  0.0001 in units of Hammer projection coordinates $x_{\rm Hammer}$ and
  $y_{\rm Hammer}$ in both directions. Contour levels are: 50, 150, 250,
  400, 600, 1000, 1500, 2000 Cepheids per 1 square degree. {\it Bottom set
  of three maps:} Maps in the Cartesian coordinates projections with the
  {\it z} axis pointing toward the SMC center. The bin size is 0.5~kpc in
  {\it x}, {\it y} and {\it z}. Contour levels on the {\it xy} plane are
  50, 200, 500, 1000, 1700, 2500, 3500, on the {\it xz} plane 50, 100, 200,
  300, 400, 500, 650 and on the {\it yz} plane 50, 150, 250, 400, 600, 800
  Cepheids per 1 kpc$^2$. White circle marks the SMC center (Graczyk \etal
  2014, Stanimiroviæ \etal 2004).}
\end{figure}

The bottom set of three maps in Fig.~13 shows Cepheid density in the
Cartesian space (see figure caption for a full description). The bottom
left map, in the {\it xy} plane, resembles the map with the Hammer
projection although the contours are more smooth and the additional
structures are not clearly visible. The projection on the {\it xz} plane
does not show any evident substructures. The densest region of the SMC is
located farther than the mean galaxy distance and falls between distances
62--70~kpc. The {\it yz} plane yields a more compelling evidence for the
existence of the northern substructure, situated in the closer part of the
SMC. Fig.~2 from Haschke \etal (2012b) shows that this substructure was not
clearly visible in the OGLE-III Cepheid data, although it somewhat emerges
in their Fig.~3.

\subsection{Ages}
We estimated ages of Cepheids in the SMC using the period--age relations
from Bono \etal (2005) for a constant metallicity $Z=0.004$.  We again
assumed that there is no metallicity gradient in the SMC, which is
supported by recent studies (Cioni 2009, Parisi \etal 2009, Deb and Singh
2014). However, some suggest that the SMC may have a low metallicity
gradient (Carrera \etal 2008, Kapakos and Hatzidimitriou 2012, Dobbie
2014), and if this was the case, it may have somewhat influenced our age
estimates. Romaniello \etal (2008) found a metallicity spread $\Delta[{\rm
Fe}/{\rm H}]\approx0.2$~dex for 12 Cepheids in this galaxy. This would
translate to a metallicity range of $Z\in(0.003,0.005)$. We made a rough
estimate by interpolating PA relations from Bono \etal (2005) and found
that such a spread in metallicity would introduce differences in age
calculations at the level of up to $\sim$10\% for first-overtone, and up to
$\approx6\%$ for fundamental mode pulsators.

\begin{figure}[htb]
\includegraphics[width=12.5cm]{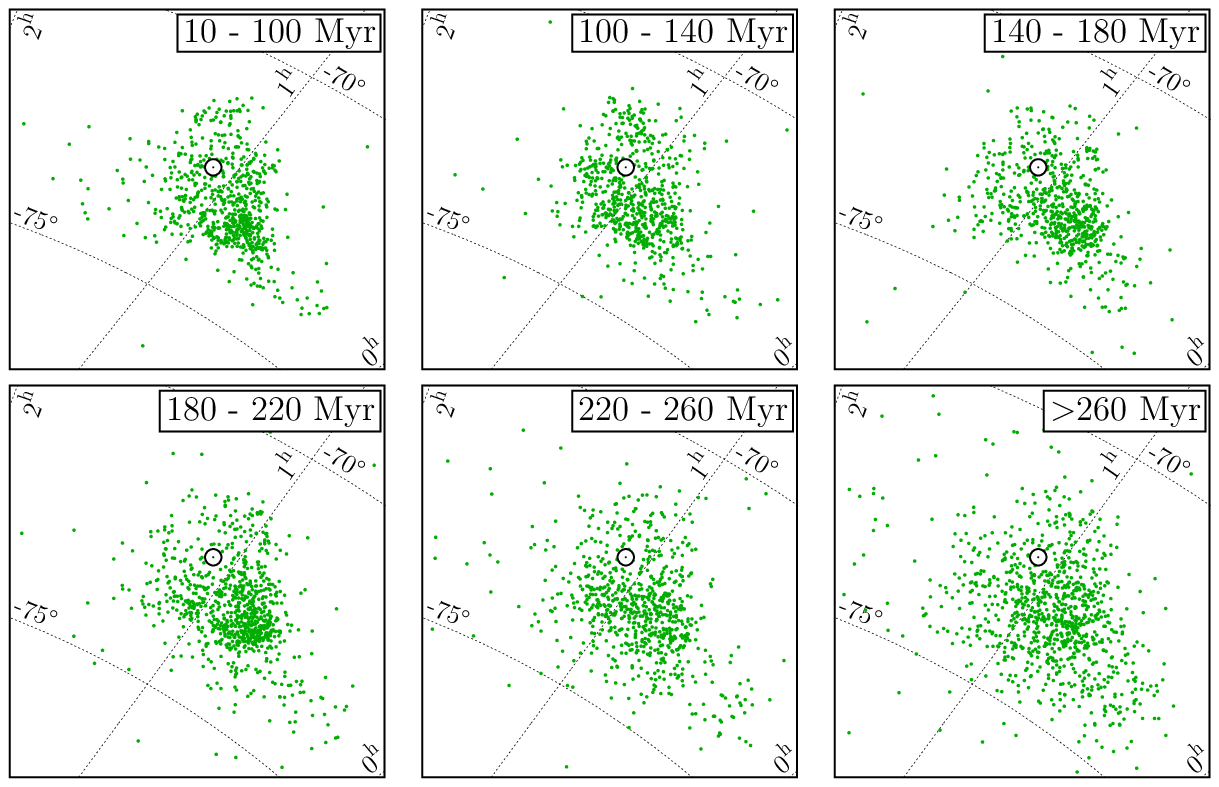}
\FigCap{Age tomography of the SMC using the relation from Bono \etal (2005)
  for a constant metallicity $Z=0.004$. The maps are in the Hammer
  projection. Note that the {\it first panel} shows an interval of 90~Myr,
  {\it last} -- 278~Myr, the {\it other ones} -- 40~Myr. White circle marks
  the SMC center (Stanimiroviæ \etal 2004).}
\end{figure}

In Fig.~14 we show the on-sky view of Cepheids in different age
intervals. The age range is larger than in the LMC, which means that the
SMC CCs population is older than that of the LMC. Young and
intermediate age Cepheids form similar structures, although young stars are
more concentrated in the north than older stars (second map in the top
row). The older the Cepheids the more they concentrate in the south-western
parts of the SMC (second map in the bottom row). The oldest stars in our
sample are rather evenly spread and do not form any obvious structures.
Our Cepheid age-tomography matches well with Fig.~13 from Rubele \etal
(2015) where the star formation rates (SFRs) for the VMC data are
shown. Recently formed stars have a ``heart-like'' structure while the
older ones are more uniformly distributed.

The differences in the distribution of younger and older stars are even
better visible in Fig.~15. The maps show Cartesian space projections and
the transformation is rotated so that the {\it z} axis is pointing toward
the SMC center. Cepheids are divided into two groups: younger than 150~Myr
and older than 150~Myr. The former group is represented with red dots and
the latter with blue dots. We clearly see that younger Cepheids are
located mainly in the closer part of this galaxy while the older ones are
distinctly farther.

\begin{figure}[htb]
\includegraphics{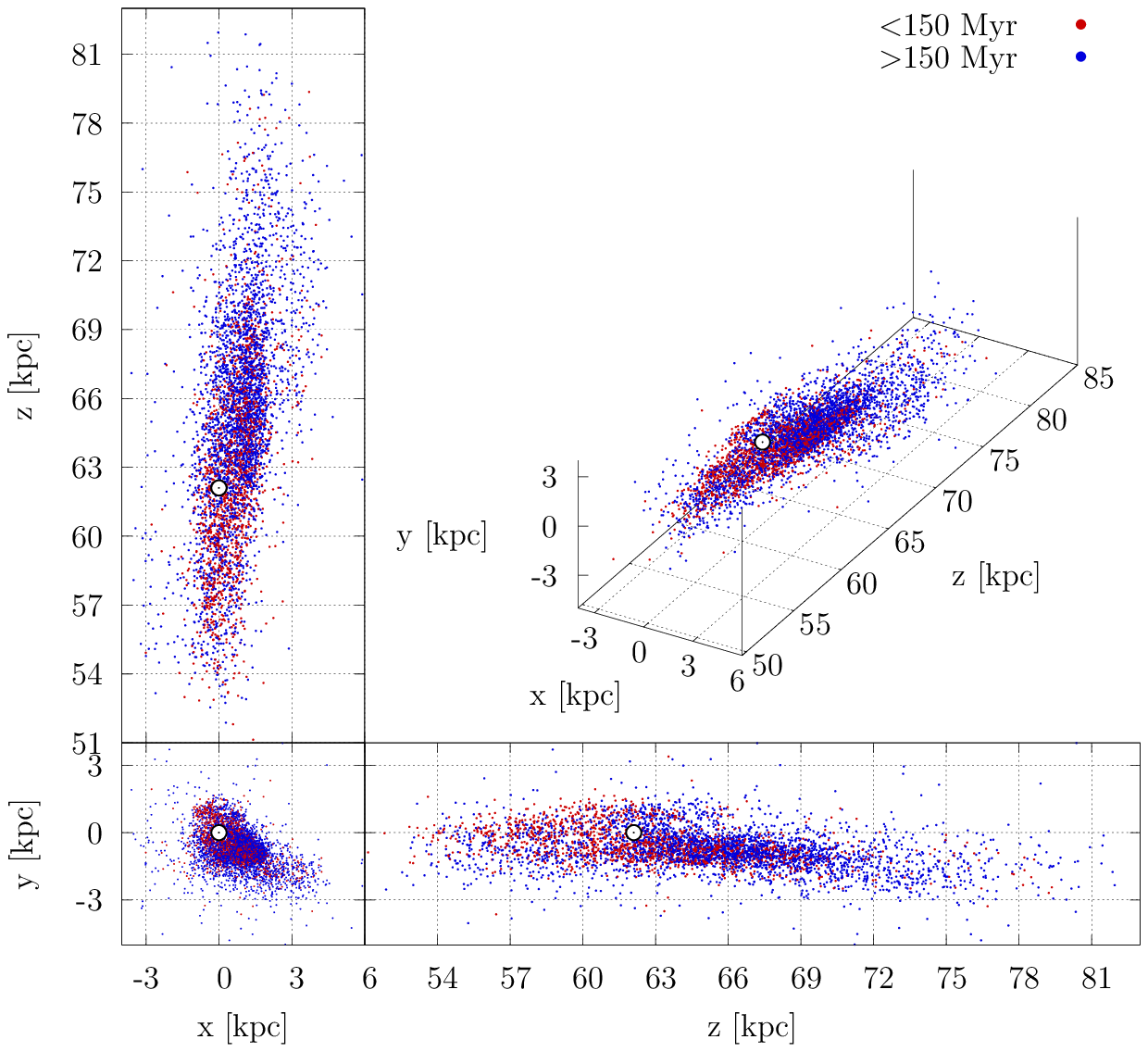}
\FigCap{Three-dimensional map of the CCs in the SMC in Cartesian
  coordinates with the {\it z} axis pointing toward the SMC center. Red
  dots represent Cepheids younger than 150~Myr and blue dots stand for
  Cepheids in the age interval of 150--300~Myr. White circle marks the SMC
  center (Graczyk \etal 2014, Stanimiroviæ \etal 2004).}
\end{figure}

\subsection{Substructures}
In order to investigate the structure of the SMC in more detail we selected
two subregions and named them south-western and northern region. The
selected areas are shown in Fig.~16. The substructures are also visible in
Fig.~12. The northern one is best visualized in the second and third top
panels and also in the first bottom panel. The south-western region emerges
in the first bottom panel and is even more clear in the following
panels. We see that the south-western region is located in the more distant
half of the SMC while the northern region is in the closer part of this
galaxy. The latter is consistent with Subramanian and Subramaniam (2012)
who stated that the north-eastern part of the SMC is located closer to us,
based on red clump and RR~Lyr stars. Both substructures are distinct on
the three-dimensional SMC maps as well as on the contour maps.

\begin{figure}[htb]
\includegraphics[width=12.5cm]{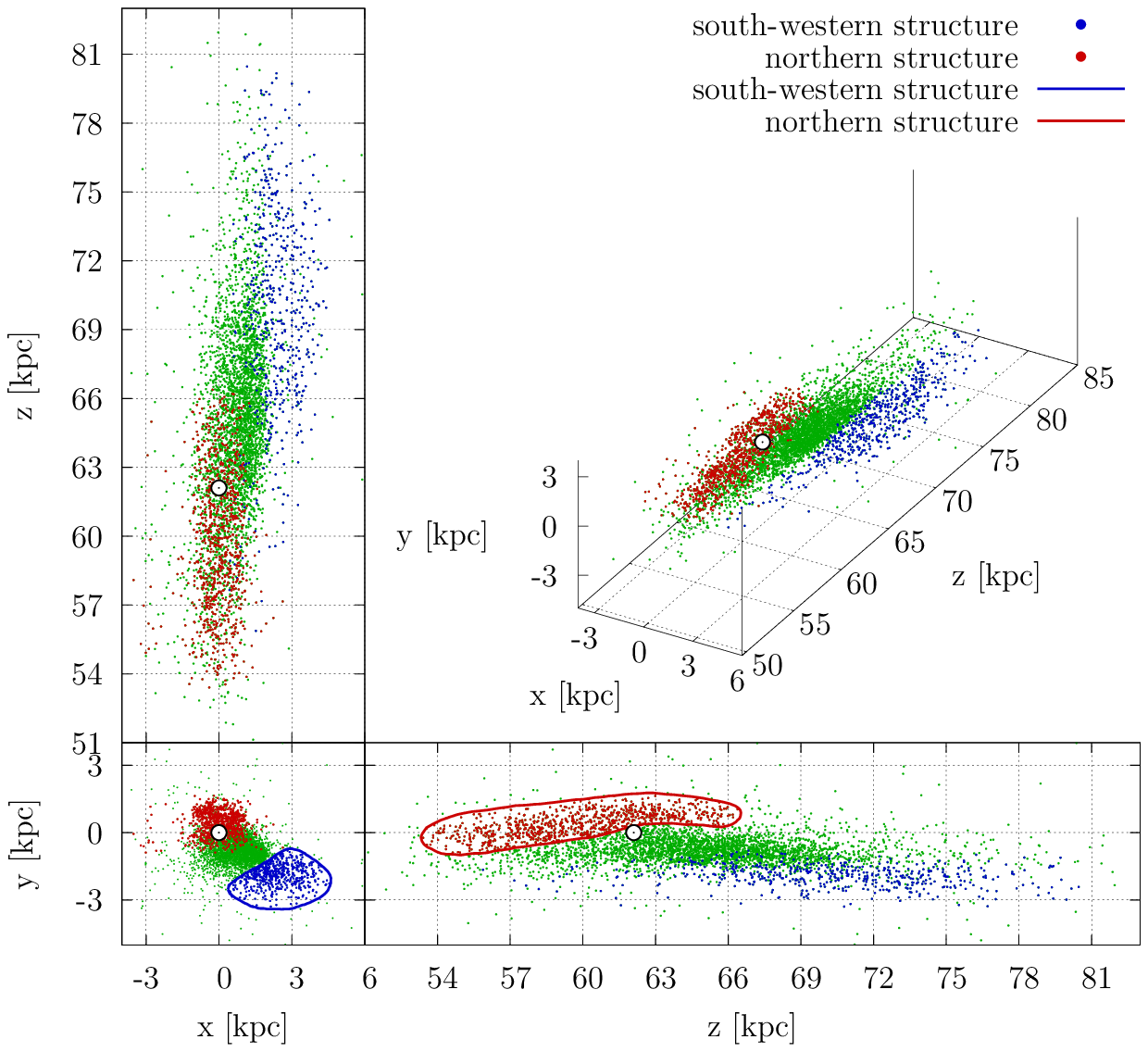}
\FigCap{Three-dimensional map of the CCs in the SMC in Cartesian
  coordinates with the {\it z} axis pointing toward the SMC center. The
  map shows selected areas for the south-western and northern regions
  marked with blue line and dots, and red line and dots,
  respectively. White circle marks the SMC center (Graczyk \etal 2014,
  Stanimiroviæ \etal 2004).}
\end{figure}

\MakeTable{|l|cc|cc|c|}{12.5cm}{Characteristics of the SMC substructures}
{\hline
\douprule
Substructure & $\langle{\rm dist\rangle}$ [kpc]& $\sigma_{\rm dist}$ [kpc]& $\langle{\rm age\rangle}$ [Myr]& $\sigma_{\rm age}$ [Myr]& N \\
\hline
\uprule
All Cepheids            & 64.62  &  4.95 &  193 & 89  &4654\\
Northern Structure      & 59.90  &  3.00 &  152 & 84  & 868\\
South-Western Structure & 70.18  &  4.44 &  233 & 88  & 525\\
\hline
\noalign{\vskip3pt}
\multicolumn{6}{p{11cm}}{Table lists median distance and age together with standard deviations, and a number of stars in each substructure.}
}

Table~8 lists median distances and ages of the SMC and its substructures,
together with standard deviations and sample numbers. Fig.~17 shows
distance and age distributions for the whole SMC as compared with its two
substructures (left panels) and with the LMC (right panels). We again see
that the south-western structure is situated in the farther half of the SMC
while the northern region is situated closer. The bottom left panel also
reveals that the latter is younger than the former and the KS test results
(Table~9) reject the hypothesis of samples coming from the same
distributions at significance level $\alpha=0.001$. This is also consistent
with our conclusions from Section~5.2, \ie that the SMC closest parts were
formed later than its more distant areas.

\begin{figure}[htb]
\includegraphics[width=12.5cm]{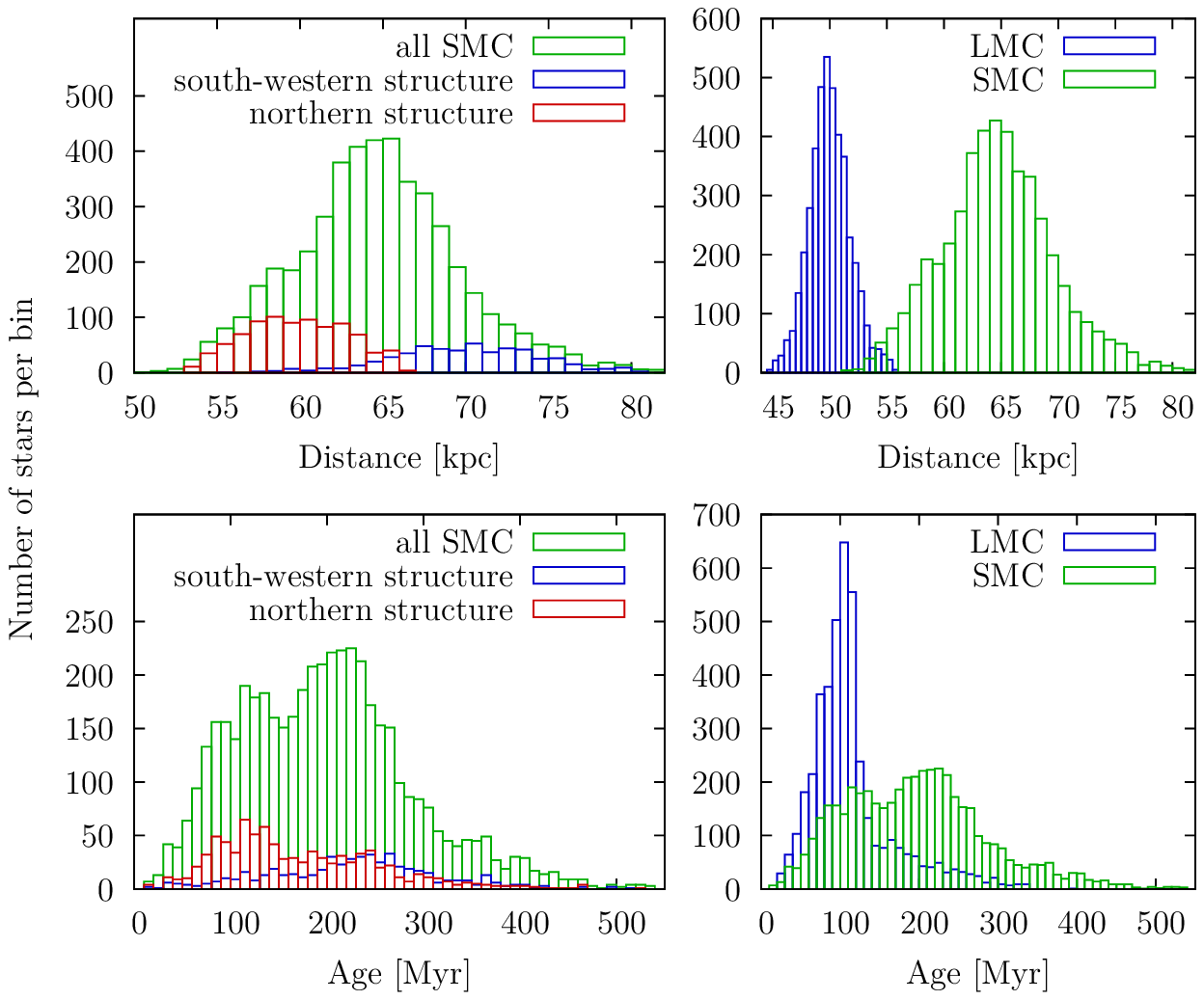}
\FigCap{Distance ({\it top row}) and age ({\it bottom row})
histograms for the SMC. {\it Left panel:} Histograms for the whole SMC Cepheid
sample (green) and separately the two selected substructures: the south-western
(blue) and northern (red) regions.
{\it Right panel:} Comparison of the SMC (green) with the LMC (blue).}
\end{figure}

\renewcommand{\arraystretch}{0.95}
\MakeTableee{|ll|ccc|ccc|}{12.5cm}{Kolmogorov-Smirnov test results in the SMC}
{\hline
\uprule
 &  & \multicolumn{3}{c|}{DISTANCE} & \multicolumn{3}{c|}{AGE} \\
\dorule
Sample 1 & Sample 2 & $D$ & $\pvalue$ & $\alpha^*$ & $D$ & $\pvalue$ & $\alpha^*$ \\
\hline
\uprule
all      & northern      &  0.471 & 0.000 & 0.001 & 0.168 & 0.000 & 0.001 \\
all      & south-western &  0.460 & 0.000 & 0.001 & 0.200 & 0.000 & 0.001 \\
northern & south-western &  0.839 & 0.000 & 0.001 & 0.334 & 0.000 & 0.001 \\
\hline
\noalign{\vskip5pt}
\multicolumn{8}{p{11cm}}{$^*\alpha$ is a significance level at which a null
hypothesis that the two samples come from the same distribution can be
rejected.}
}

The right panels illustrate differences between the LMC and the SMC. The
top panel shows that both galaxies have Gaussian-like distance
distributions although the SMC has a bump on the left side of the
maximum. The age histogram in the bottom panel shows that the LMC Cepheids
are on average significantly younger than the SMC objects. The oldest LMC
Cepheids are $\approx 390$~Myr old, while the oldest SMC stars are $\approx
540$~Myr old.

The SMC must have had two epochs of star formation. It is reflected in its
bimodal Cepheid age distribution. The younger bump has its maximum close to
the LMC peak ages (around 110~Myr) while the second bump is at the age of
about 220~Myr. The two SMC peaks are separated by the local minimum at
about 150~Myr. Fig.~15 shows differences in their spatial distribution, \ie
the youngest Cepheids are closer to us than the older ones.

The two-peak nature of the age distribution in the SMC was also detected by
Subramanian and Subramaniam (2015). Their Fig.~9 is very similar to ours in
the context of the maxima, the peak separation and the age range. The
spatial distribution of different-age Cepheids is consistent in both
studies (see Fig.~10 from Subramanian and Subramaniam 2015 and Fig.~14 in
this work), even though Subramanian and Subramaniam (2015) used the
period--age--color relations from Bono \etal (2005) for dereddened data,
while we used the simpler period--age relation.

On the other hand, there is only one episode of extensive Cepheid formation
in the LMC, coincident with the younger SMC bump, followed by a slow
decline toward older ages. This shows that Clouds had a different Cepheid
formation history, possibly with a common episode. At the same time it does
not mean that the Clouds had a different SFH, since we only concentrate on
CCs in this paper. Moreover, because the SMC has lower metallicity than
LMC, the Cepheids in the former galaxy may be more massive and thus older.

\Section{The Magellanic Bridge}
From our initial sample of Cepheids in the Magellanic System we decided to
classify nine as the Magellanic Bridge objects. Their parameters are listed
in Table~10. We provide Cepheids' ID from the OGLE Collection of Variable
Stars along with the local ID that we use in this work (M1,$\dotsc$,M9),
pulsation period $P$, {\it I}- and {\it V}-band magnitudes, equatorial
coordinates for epoch J2000.0, distance $d$ and estimated age. The distance
uncertainty does not include the mean LMC distance uncertainty (from
Pietrzyñski \etal 2013 $d_{\rm LMC}=49.97\pm0.19\ {\rm
  (statistical)}\pm1.11\ {\rm (systematic)}\ {\rm kpc}$). The list contains
four fundamental-mode Cepheids, four first-overtone pulsators and one
double-mode oscillator (1O2O) for which we analyzed its lowest mode (1O).

\renewcommand{\TableFont}{\scriptsize}
\MakeTable{c!{\color{black}\vrule}c@{\hspace{4pt}}r@{.}lc@{\hspace{4pt}}c@{\hspace{4pt}}c@{\hspace{7pt}}c@{\hspace{9pt}}c@{\hspace{4pt}}c}{12.5cm}{Magellanic Bridge Cepheids}
{\hline
\uprule \multirow{2}{*}{P. mode} & \multicolumn{6}{l}{OCVS ID} \\
\dorule & Loc. ID & \multicolumn{2}{c}{$P$ [d]} & $I$ [mag] & $V$ [mag]& RA & Dec & $d~[{\rm kpc}]^{(a)}$ & Age [Myr] \\ 
\hline
\uprule \multirow{8}{*}{F} & \multicolumn{4}{l}{OGLE-SMC-CEP-4956}                                           \\
\dorule & M1 & 1&1162345  & 17.372 & 17.930   & $3\uph23\upm24\zdot\ups90$ & $-74\arcd58\arcm07\zdot\arcs3$  & $72.11\pm2.01$ & $283\pm58$ \\ 
\arrayrulecolor{lightgray}\cline{2-10}\uprule & \multicolumn{4}{l}{OGLE-SMC-CEP-4957}                        \\
\dorule & M2 &  1&4300017 & 17.376 & 18.112   & $3\uph43\upm04\zdot\ups54$ & $-76\arcd56\arcm02\zdot\arcs6$  & $74.61\pm2.08$ & $232\pm48$ \\ 
\arrayrulecolor{lightgray}\cline{2-10}\uprule & \multicolumn{4}{l}{OGLE-LMC-CEP-3376$^{(\star)}$}            \\
\dorule & M3 &  1&1589986 & 15.892 & 16.350   & $4\uph01\upm38\zdot\ups02$ & $-69\arcd28\arcm40\zdot\arcs5$  & $40.13\pm1.12$ & $275\pm56$ \\
\arrayrulecolor{lightgray}\cline{2-10}\uprule & \multicolumn{4}{l}{OGLE-SMC-CEP-4953$^{(\star)(b)}$}         \\
\dorule & M4 & 21&3856946 & 12.967 & 13.821   & $2\uph20\upm49\zdot\ups46$ & $-73\arcd05\arcm08\zdot\arcs3$  & $53.93\pm1.50$ & $27\pm5$ \\
\arrayrulecolor{black}\hline\uprule\multirow{8}{*}{1O}  & \multicolumn{4}{l}{OGLE-SMC-CEP-4955$^{(\star)}$}  \\
\dorule & M5 &  2&0308924 & 15.675 & 16.281   & $2\uph42\upm28\zdot\ups88$ & $-74\arcd43\arcm17\zdot\arcs6$  & $60.04\pm1.65$ & $120\pm19$ \\ 
\arrayrulecolor{lightgray}\cline{2-10}\uprule & \multicolumn{4}{l}{OGLE-LMC-CEP-3377$^{(\star)}$}            \\
\dorule & M6 &  3&2144344 & 14.629 & 15.291   & $4\uph04\upm28\zdot\ups88$ & $-75\arcd04\arcm47\zdot\arcs1$  & $48.76\pm1.34$ & $73\pm12$ \\
\arrayrulecolor{lightgray}\cline{2-10}\uprule & \multicolumn{4}{l}{OGLE-SMC-CEP-4952}                        \\
\dorule & M7 &  1&6414839 & 16.901 & 17.535   & $2\uph04\upm09\zdot\ups38$ & $-77\arcd04\arcm38\zdot\arcs4$  & $89.51\pm2.46$ & $151\pm24$ \\ 
\arrayrulecolor{lightgray}\cline{2-10}\uprule & \multicolumn{4}{l}{OGLE-SMC-CEP-4954}                        \\
\dorule & M8 &  0&8883941 & 17.156 & 17.512   & $2\uph21\upm28\zdot\ups45$ & $-65\arcd45\arcm22\zdot\arcs4$  & $80.71\pm2.22$ & $291\pm47$ \\
\arrayrulecolor{black}\hline\uprule\multirow{2}{*}{1O2O} & \multicolumn{4}{l}{OGLE-SMC-CEP-4951$^{(\star)}$} \\
\dorule & M9 &  0&7170500 & 16.769 & 17.222   & $2\uph02\upm33\zdot\ups88$ & $-75\arcd30\arcm48\zdot\arcs0$  & $54.44\pm1.50$ & $366\pm59$ \\
\arrayrulecolor{black}\hline
\noalign{\vskip5pt}
\multicolumn{10}{p{12.5cm}}{$^{(\star)}$These stars form a continuous-like
  connection between the Magellanic Clouds. $^{(a)}$~The distance
  uncertainty does not include the mean LMC distance uncertainty from
  Pietrzyñski \etal (2013) $d_{\rm LMC}=49.97\pm0.19\ {\rm
    (statistical)}\pm1.11\ {\rm (systematic)}\ {\rm kpc}$. $^{(b)}$~The
  OGLE-IV Collection of Classical Cepheids provides only the {\it V}-band
  magnitude for this object. The star is saturated on the standard OGLE-IV
  {\it I}-band reference image. Presented here {\it I}-band magnitude and
  more accurate period determination comes from dedicated re-reduction of
  the OGLE images.}}

Soszyñski \etal (2015) classified five Cepheids as MBR objects. Our Bridge
sample contains four more objects than their sample, which is not
surprising, as our classification was based not only on the on-sky
projected locations of the Cepheids (see Fig.~18), but also on their
three-dimensional distribution (see Fig.~19). Even so one can argue about
the classification of M9 Cepheid. This object is close to the whole SMC
sample and could be assigned to the SMC Wing. Nevertheless, we believe that
this object is connecting the SMC Wing with the Bridge and may as well be
classified as a Bridge Cepheid.

\begin{figure}[htb]
\includegraphics[width=12.9cm]{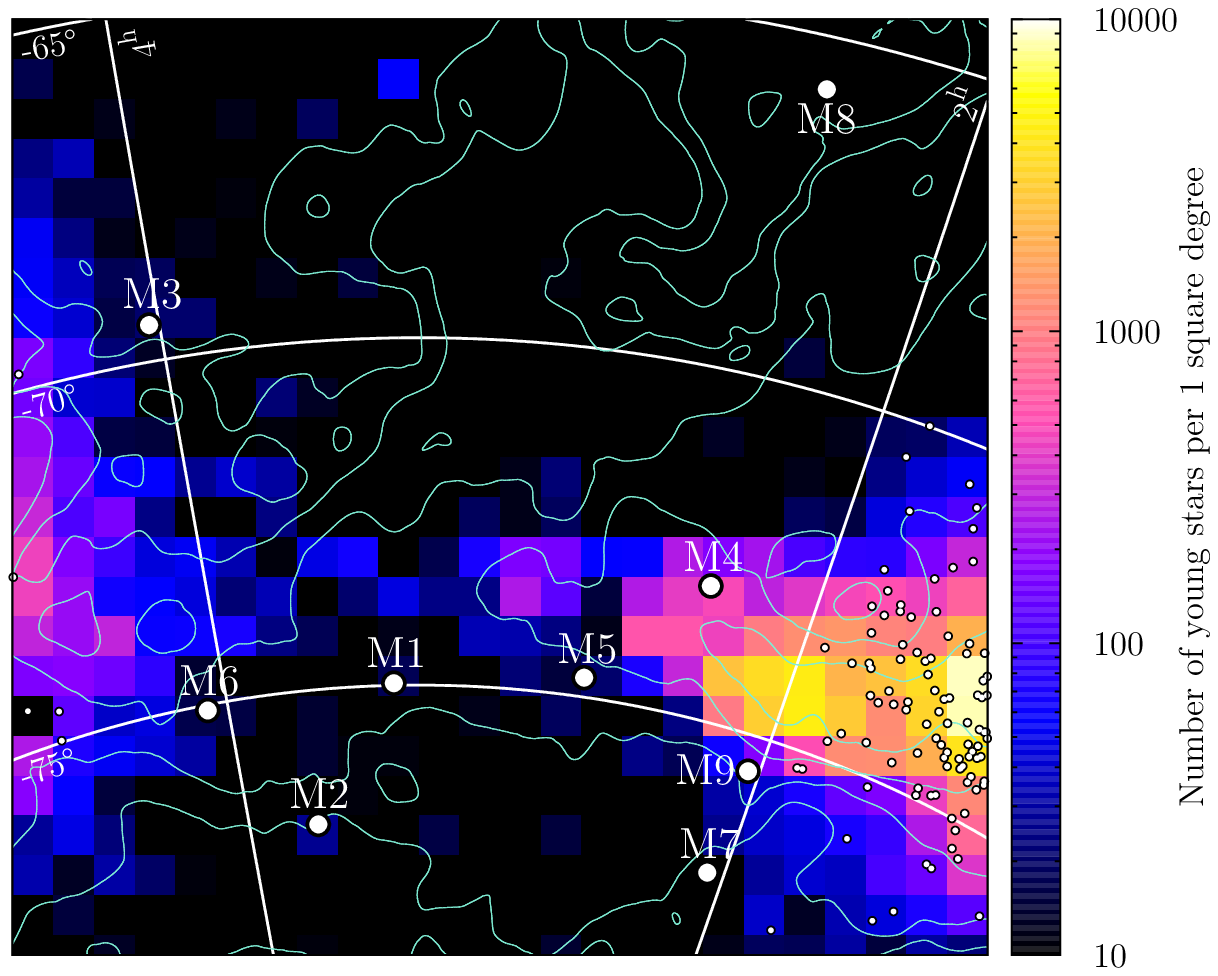}
\FigCap{CCs in the Magellanic Bridge area over the spatial
  density map of the Young Population stars from Skowron \etal
  (2014). Labels M1--M9 represent the Cepheids' local IDs from Table~7. The
  map is represented in a Hammer equal-area projection centered at
  $\alpha_{\rm cen}=3\uph18\upm$, $\delta_{\rm cen}=-70\arcd$. The
  color-coded value of each ``pixel'' is a logarithm of the number of stars
  per square degree area, while each ``pixel'' area is $\approx0.335$ square
  degrees. Light green contours mark neutral hydrogen (HI) emission
  integrated over the velocity range $80<\vv<400$~km/s, where each
  contour represents the HI~column density twice as large as the
  neighboring contour. HI~column densities are in the range
  $10^{20}-4\times10^{21}~{\rm cm}^{-2}$. Data were taken from the LAB
  survey of Galactic HI~(Kalberla \etal 2005).}
\end{figure}

Fig.~18 shows the location of our Cepheids with respect to the HI density
contours (Kalberla \etal 2005) and the young stellar population discovered
by Skowron \etal (2014). Almost all Cepheids' locations are well correlated
with the HI contours and with the young stellar population space density
distribution. Especially M4, which is also the youngest Cepheid in our MBR
sample, is located in one of the densest young population regions from
Skowron \etal (2014) near the SMC.

Skowron \etal (2014) showed that there exists a continuous connection
between the two Magellanic Clouds built up of the young stars (age
$<1$~Gyr). The on-sky distribution of Bridge Cepheids also forms a
continuous connection and adds to the overall distribution of the young
population. These are Cepheids named M6, M1, M5, M4, M9 (see Fig.~18). If
we look at their three-dimensional distribution in Fig.~19, they fall along
a line between the Clouds in the {\it xy} plane. The {\it xz} and {\it yz}
planes show that M6, M5, M4 and M9 indeed form a connection between the
Clouds. M3 may also contribute to this structure. On the other hand, M1 and
M2 lie significantly farther. Moreover, they are located in the outskirts
of the young population density contours from Skowron \etal (2014) which
may indicate their different origin. Similarly, M7 and M8 are located even
farther from both Clouds and also far from the young population density
contours, thus they do not belong to the genuine Bridge population. These
two Cepheids may contribute to the Counter Bridge predicted in numerical
simulations (Diaz and Bekki 2012).  We discuss this in details in
Section~7.

\begin{figure}[htb]
\includegraphics[width=12.9cm]{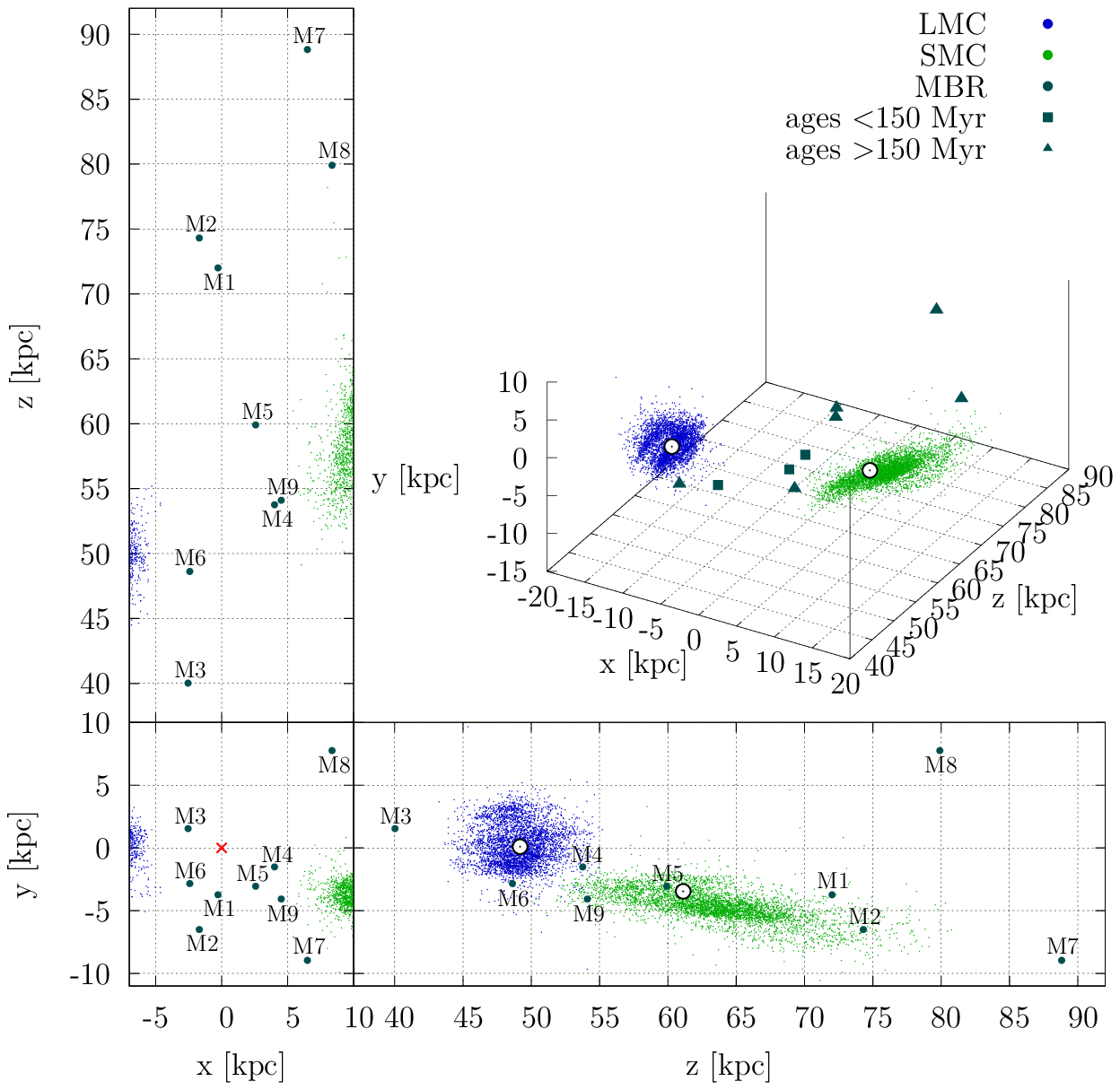}
\FigCap{Three-dimensional map of CCs in the Magellanic Bridge in Cartesian
  coordinates with the {\it z} axis pointing toward $\alpha_{\rm
    cen}=3\uph20\upm$, $\delta_{\rm cen}=-72\arcd$. Blue dots represent the
  LMC, green dots SMC and the large dark teal dots -- the MBR. Labels
  M1--M9 represent the Cepheids' local IDs from Table~7. Red cross stands
  for the observer's location. White circle marks the LMC (Pietrzyñski
  \etal 2013, van der Marel and Kallivayalil 2014) and SMC (Graczyk \etal
  2014, Stanimiroviæ \etal 2004) centers.}
\end{figure}

The Cepheids in the Magellanic Bridge are very spread along the line of
sight. The closest star (M3) is located at $d\approx40$~kpc thus it is closer
to us than any LMC Cepheid. The farthest (M7) is at almost 90~kpc and this
is farther than any SMC Cepheid. This again shows that not all MBR Cepheids
form a continuous connection between the Clouds, and rises a question about
their origin and how they got to their current location. On the other hand,
we do observe stars located far from the LMC and SMC all around these
galaxies (\ie see the LMC Cepheid at $\alpha\approx5\uph30\upm$,
$\delta\approx-56\arcd$ or SMC Cepheid at $\alpha\approx23\uph30\upm$,
$\delta\approx-68\arcd$ in Fig.~3). These objects were probably ejected
from the galaxies in random directions. Some of our MBR Cepheids may belong
to the outliers population.

The ages of Bridge Cepheids were again calculated using the period--age
relation from Bono \etal (2005). There are different relations for
different metallicities. In the case of the Bridge the gas metallicity is
about $Z_{{\rm MBR}}\approx0.1\ Z_{\odot}$ (Lehner \etal 2008) or slightly
higher ($-0.5<\log(Z_{{\rm MBR}}/Z_{\odot})<-1$ from Misawa \etal 2009,
although this was measured along the line of sight in an area that is
possibly not mixed with metal-poor gas, as it is in other regions of the
Bridge). Moreover, the 0.1 solar metallicity in the MBR is consistent with
the Magellanic Stream metallicity (Fox \etal 2013). Thus we can assume
$Z_{\rm MBR}=0.002$ for the Bridge Cepheids (if we first assume that they
were formed {\it in-situ}). Note that Bono \etal (2005) do not provide the
period--age relation for this metallicity -- the lowest value is $Z=0.004$
(typical for the SMC). We therefore use this relation for the MBR Cepheids,
keeping in mind that it is just a rough estimate.

The youngest Cepheid is M4 and its age estimate is 27~Myr. Its location is
well correlated with the young population density contours from Skowron
\etal (2014). This star was probably formed together with other young stars
in the Bridge. Another young Cepheid is M6 and its age estimate is
74~Myr. This star is located at a distance close to the mean LMC distance
and is $\approx7.1$~kpc from the center of the LMC, which is much farther
than any other LMC Cepheid. The oldest Cepheid is M9 and it is
$\approx370$~Myr old. This star is located fairly close to the SMC Wing and
may be classified as the Wing object.  Two Cepheids are aged between
100--200~Myr. One of them is the farthest one -- M7, which is about 150~Myr
old. The other four Cepheids are in the ages range 200--300~Myr. One of
them is the closest object, the other three are located at distances
72--81~kpc.

\Section{Discussion}
\subsection{Three-Dimensional Structure and Substructures: the LMC}
The LMC has a bar that is thought to be offset from the center of this
galaxy by about 0.5~kpc. First suggestions that the bar may not be aligned
with the disk plane were based on the microlensing events (Zhao and Evans
2000). The offset of about 0.5~kpc was measured and used in many studies
(\eg Nikolaev \etal 2004, Subramanian and Subramaniam 2013, van der Marel
and Kallivayalil 2014). The offset had also been predicted by numerical
models, \eg Bekki (2009) concludes that it is not the bar that is offset
but the entire disk population. Besla \etal (2012) had reproduced not only
the off-center bar but also the spiral structure of the LMC with one arm.

In this work we redefine the idea of the LMC bar. By examining the distance
and age distributions of the central parts of the LMC we argue that the bar
comprises of not only the central-eastern region considered to be the
``classical'' bar, but also of the western region, as shown in Fig.~7. In
the distance and age regime both parts are continuously connected, making
the homogeneous, though asymmetrical, structure. The redefinition of the
bar moves the dynamical center of the LMC to the center of the bar.

The mean distance of the redefined bar is close to the mean LMC distance
and we do not observe any significant offset. That is not consistent with
the value of 0.5~kpc from the literature, and the reason may be a different
definition of the bar region (see Fig.~14 in Nikolaev \etal 2004, Fig.~2 in
Subramaniam and Subramanian 2009 and Figs.~1 and 7 in Haschke \etal 2012a
-- bar areas are consistent with our eastern part of the bar from Fig.~7).
However, if we use the ``classical'' LMC bar, we still do not see a
significant offset from the galaxy center (the ``classical'' bar is located
closer to us by only $\approx0.07$~kpc), contrary to the cited studies.

We fitted a plane to the entire Cepheid population in the LMC as well as to
its substructures. The whole LMC sample shows no offset along the line of
sight as compared to the mean LMC distance from Pietrzyñski \etal (2013)
and that is expected from a correct fitting procedure. The obtained
inclination and position angles are consistent with values from the
literature (see Table~7). The {\it rms} of our fit is about 1.5~kpc, which
is partly a ``natural spread'' of the method described in Section~3, and
partly a contribution of the extra-planar features of the LMC. Nikolaev
\etal (2004) found that the disk is warped, with a distortion amplitude
$\gtrsim0.3$~kpc. This warp explains high $\chi^2/{\rm dof}$ values for
planar disk in our fits. On the other hand, Subramanian and Subramaniam
(2013) found that the disk can be divided into two differently inclined
parts -- the inner and the outer -- separated at the radial distance from
the LMC center of $3\arcd$. The inner disk would be more warped than the
outer. They also concluded that the bar is offset but is still a co-planar
feature. They classified structures as extra-planar if their deviation is
$>1.5$~kpc. Olsen and Salyk (2002) had previously identified warps in the
disk in similar locations. The detailed modeling of the extraplanar
features of the LMC disk is beyond the scope of this paper.

We also fitted a plane to Cepheids in the redefined bar and found a small
offset of about $-0.09$~kpc from the mean LMC distance which is
statistically insignificant within $3\sigma$ uncertainty. We are aware that
fitting a plane to the bar is not the best approach because of the
nonplanar nature of this structure. Nikolaev \etal (2004) suggest caution
when deriving parameters such as viewing angles for the inner LMC
structures. Also Subramanian and Subramaniam (2013) stated that the
structure of the bar is not smooth and some of its parts are located closer
to us than other.

Interestingly, when fitting a plane to the northern arm sample we found an
offset of about $-0.48$~kpc. This means that the arm is located closer to
us than the whole LMC. Moreover, the arm lies in a different plane (has
different inclination and position angle) than the whole LMC sample and
this result is statistically significant.

The OGLE-IV CCs data set clearly shows the bar and the main northern arm of
the LMC. We also tried to localize less prominent structures in other parts
of this galaxy. In the north we identified an additional small spiral arm
(NA2, see Section~4.3). This finding is consistent with the latest results
from Besla \etal (2016) who analyzed deep optical images of the LMC and
identified multiple spiral arms. Both structures are at precisely the same
location -- compare our northern arm 2 in Fig.~7 with multiple spiral arms
in Fig.~3 from Besla \etal (2016).  The structures that we see in the
southern part of the LMC are not as prominent and do not form a spiral arm,
which is also consistent with conclusions from Besla \etal (2016). However,
it does not exclude the possibility that there exists a sparse spiral arm
connected with the south-eastern part of the bar which is not clearly
visible in the CCs distribution.

We compare our results to those obtained by Haschke \etal (2012a) from the
OGLE-III Cepheid data. What is striking -- the distances they derived are
substantially larger than ours. Cepheid distances fall in the range of
44--56~kpc in this work, and 45--60~kpc in the work of Haschke \etal
(2012a). This discrepancy is also reflected in their mean LMC distance of
$53.9\pm1.8$~kpc which is not consistent with the literature (as
highlighted by de Grijs \etal 2014). The method of determining distances
was similar in both studies, but we used a reddening-free Wesenheit index
and determined distances relative to the most accurate LMC distance
measurement (Pietrzyñski \etal 2013), while Haschke \etal (2012a)
calculated absolute distances based on the {\it I}- and {\it V}-band
magnitudes corrected for extinction. Thus the problem could lie in the
dereddening method or the reddening maps used, as also suggested by de
Grijs \etal (2014).

It is also worth noting that the OGLE-III collection of Cepheids in the LMC
used by Haschke \etal (2012a) did not include the northern arm and some of
the southern parts of this galaxy. For comparison see the lower-right panel
in Fig.~12 of Moretti \etal (2014) where they compare the OGLE-III CCs with
the EROS-2 data. Nevertheless, the results that did not include the
northern arm should also be consistent with ours, since the northern arm is
closer to us than the rest of the galaxy, while the southern parts are at
approximately the same distance.

\subsection{Three-Dimensional Structure and Substructures: the SMC}
We find that the SMC is extremely elongated almost along the line of
sight. Its size along the {\it z} Cartesian axis is about 4--5 times
larger than along the {\it x} and {\it y} axes. This is consistent with the
latest structural analysis of the SMC performed by Scowcroft \etal (2016),
based on mid-infrared Spitzer data for 92 Cepheids. The comparison of
Fig.~6 in Scowcroft \etal (2016) with our Fig.~15 or 16 shows a similar
spread along each of the axes, although the substructures are only visible
in the OGLE-IV data, as the sample is about 50 times more numerous.

We agree with Scowcroft \etal (2016) that the standard parameters such as
the inclination and position angle are not adequate for describing a galaxy
with such an elongated shape, even though such parameters were determined
in many studies (\eg Stanimiroviæ \etal 2004, Subramanian and Subramaniam
2012, Haschke \etal 2012b, Subramanian and Subramaniam 2015). Scowcroft
\etal (2016) claim that the shape of the SMC can be best characterized as a
cylinder. We would rather describe it as a tri-axial elongated ellipsoid,
although the existence of the ``off-axis'' structures makes it even more
complicated and separate fits for the main body and the substructures might
be necessary (see Fig.~16).

We would expect our results to be coherent with those of Haschke \etal
(2012b), based on the OGLE-III CCs catalog, as the number of Cepheids is
similar and the main body of the SMC is clearly visible in both data sets
(Fig.~1 in Haschke \etal 2012b and Fig.~13 in this work). Any differences in
conclusions would be a result of different methods of distance
determinations, as noted in Section~7.1. They obtained the median distance
to the SMC for the Cepheid sample of $63.1\pm3.1$~kpc which is consistent
with the literature (de Grijs and Bono 2015) and with the median SMC
distance of $64.6\pm4.9$~kpc derived from our sample.

However, the bottom map in Fig.~3 of Haschke \etal (2012b) suggests that
the SMC is not very elongated along the line of sight and rather has a
disk-like structure, although the spread in distances of about 30~kpc is
consistent with our results, so it is only an effect of the chosen
projection.  The difference is in the distance range, which is about 50~kpc
to 80~kpc in this study, and 45~kpc to 75~kpc in Fig.~5 of Haschke \etal
(2012b).

We also compare our results with those of Subramanian and Subramaniam
(2015), who analyzed Cepheids from the OGLE-III catalog. Their Fig.~7 shows
similar SMC geometry as our Fig.~15, although one has to keep in mind that
the {\it x} and {\it y} are swapped with respect to our plots, and the
resolution is different for each of their axes, which gives a false
impression about the shape of this galaxy. Fig.~6 of Subramanian and
Subramaniam (2015) shows the fitted plane along the axis of the steepest
gradient and the {\it z} axis. Note that here the scale of the {\it z} axis
is 10 times smaller than the scale of the axis of the steepest gradient,
thus rising a question about the relevance of such fit. The gradient they
observe is rather an effect of the northern substructure being closer to us
(see Fig.~16 in this paper), than the SMC having an inclined plane in the
{\it xy} projection.

Subramanian and Subramaniam (2015) also detected some extra-planar features
in their sample, under the assumption that there is an actual SMC plane.
We do not support this scenario, as we show that there is no SMC plane as
such, and the galaxy can be described as a tri-axial ellipsoid, elongated
along the {\it z} axis. In this case, the reported extra-planar features
would simply be parts of the main body of the SMC or one of the
substructures shown in Fig.~16.
\vspace*{5pt}
\subsection{LMC-SMC Interactions and the Magellanic Bridge}
\vspace*{7pt}
The OGLE-IV Cepheid data show that the Magellanic Clouds are rotated
toward each other (see Fig.~2). In fact, the closest SMC Cepheids are at
similar distances as the farthest LMC objects in our sample. Moreover, the
Clouds' closest on-sky locations are also the closest in the sense of
distances and three-dimensional distribution. That is perfectly consistent
with Scowcroft \etal (2016).

The collision model by Besla \etal (2012) predicts that the Clouds had a
close interaction about 200--300~Myr ago (see Gardiner \etal 1994, Gardiner
and Noguchi 1996, R\r{u}\v{z}i\v{c}ka \etal 2010, Diaz and Bekki
2012). Both galaxies should have trails due to such interaction. It is also
possible that the co-rotation of the Magellanic Clouds has the same origin
(Scowcroft \etal 2016). Fig.~10 in Scowcroft \etal (2016) shows the
predicted SMC spheroid distribution (a model by Diaz and Bekki 2012) along
with the analyzed Cepheids. We compare it to our {\it xz} projection in
Fig.~16 where the {\it z} axis is along the distance and the {\it x} axis
-- along the right ascension (for this comparison see Fig.~6 in Scowcroft
\etal 2016). We see that our Cepheids extend even farther but still along
the gradient predicted by the model.

A model by Besla \etal (2012) predicts that there should exist a stellar
counterpart to the gaseous Magellanic Bridge, in the area between the
Clouds.  It should mainly consist of a young population of stars formed
{\it in-situ}. Such young stars were already observed in the MBR (Irwin
\etal 1985, Demers and Battinelli 1998, Harris 2007, N\"oel \etal 2013,
2015, Skowron \etal 2014), as well as intermediate-age stars (N\"oel \etal
2013, 2015) and older population candidates (Bagheri \etal 2013). Moreover,
Skowron \etal (2014) showed that there is a continuous connection between
the two Clouds made of young stars (ages $<1$~Gyr). According to Besla
\etal (2012), the stars in the Bridge should follow the Clouds past
trajectories.

In Fig.~18 we compared the OGLE-IV Cepheid locations in the Bridge with the
young stellar stream from Skowron \etal (2014). The on-sky locations are
well correlated -- most of the Bridge Cepheids are situated within the
contours of young population column densities. However, Fig.~19 shows that
only five of nine stars from our sample form a coherent structure in
three-dimensions. This raises questions about origin of the other four
Cepheids and makes an important constraint for numerical models of the
Magellanic Clouds interactions. On the other hand, these Cepheids may be
the LMC or SMC outliers ejected from these galaxies in random directions
that we now observe in the Bridge area.

Moreover, ages of our Bridge Cepheids are compatible with the assumption
that the Bridge was created during the last interaction of the Clouds (\eg
Gardiner \etal 1994, Gardiner and Noguchi 1996, R\r{u}\v{z}i\v{c}ka \etal
2010, Diaz and Bekki 2012, Besla \etal 2012). Models predict that this
interaction happened 200--300~Myr ago and most of our Cepheids are younger
than that. This indicates that they were formed outside of the Clouds -- in
the Bridge.

Diaz and Bekki (2012) model predicts not only the existence of the
Magellanic Bridge but also another structure, that they named the Counter
Bridge. It is a tidal feature of the same origin as the ``classical''
Bridge. The model reveals it as a dense and clearly defined stream that
extends away from the SMC up to the distances of about 85~kpc. Authors
conclude that the location of the Counter Bridge may cause higher levels of
optical depth in the SMC and especially in its north-eastern parts. Because
of the significant SMC elongation along the line of sight, the farthest
stars belonging to the SMC population may be mixed with the unbound stars
that should be properly classified as Counter Bridge objects.

Nidever \etal (2013) discovered a distance bimodality in the eastern SMC
using red clump stars, but mean distances of both components were too low
to be a stellar counterpart of the Counter Bridge, although the authors
argue, that the closer structure located in front of the main SMC body
forms a connection between the Magellanic Bridge and the SMC.

Subramanian and Subramaniam (2015) claim to have detected the stellar
counterpart of the Counter Bridge. They have classified it based on the
fitted plane and the extra-planar structures that they discovered in front
of as well as behind the plane (see Figs.~7 and 14 in Subramanian and
Subramaniam 2015). As we previously argued, the plane fitting in the case
of the SMC is illegitimate, making the claims about the stellar part of the
Counter Bridge an overstatement.

However, if the Counter Bridge was visible in the OGLE-III data set
(analyzed by Subramanian and Subramaniam 2015) it should also be detectable
in our sample. Fig.~2 shows all the fundamental-mode and first-overtone CCs
from the OGLE Collection of Variable Stars, many of which are much farther
(or closer) than the mean SMC distance, and these are marked with gray
dots. These stars were classified as outliers from the P-L relation and
removed from our sample in further analysis. While most of them are blends,
we cannot rule out the possibility that some of these stars may by
candidates for the Counter Bridge population (distances $>80$~kpc),
especially that two genuine Bridge Cepheids are located near or farther
than 80~kpc.

Diaz and Bekki (2012) concluded that the Counter Bridge stars may mix with the
SMC population. If this is the case, then it is possible that we observe the
Counter Bridge as a stellar structure but we are unable to separate it from
the SMC sample.

\subsection{Ages}
Indu and Subramaniam (2011) suggested that the LMC perigalactic passage
about 200~Myr ago pulled out the HI to the north of this galaxy.  Because
of the LMC's motion through the Galaxy halo the star-forming processes
began. One of the SFR peaks that they detected is at about 90--100~Myr,
which coincides with the age peak for the LMC Cepheids in our sample at
104~Myr.  Harris and Zaritsky (2009) also detected a peak in the age
distribution in the LMC around 100~Myr, although there are different maxima
in different parts of this galaxy -- the SFH of the LMC is not uniform. The
peak at about 100~Myr is observed mainly in the bar, and this is consistent
with our results, as most of the Cepheids are located in the bar. On the
other hand, Joshi and Joshi (2014) detected an intensified SF episode about
125--250~Myr ago, which is slightly older than 100 Myr found in this
analysis, but is still consistent within errors. The difference is most
probably due to different PA relations used.

The bottom right panel of Fig.~17 also shows that the younger age peak of
the SMC at about 120~Myr correlates with the LMC peak, which suggests a
common SF episode. This result is consistent with Inno \etal (2015) who
discovered that the Clouds had an active SFH during the last 400 Myr and
that there are age distribution similarities between the two galaxies.
Another common SF maxima in the Clouds were already seen at 500~Myr and
2~Gyr (Harris and Zaritsky 2009).

In the case of the the SMC, Indu and Subramaniam (2011) detected the shift
in the center of the population of stars younger than 500~Myr in the
north-east direction. That is the direction toward the LMC. We also
noticed that younger stars from our sample tend to clump in the north. The
authors also showed that the rate of this shift changed at 200~Myr and was
faster from that time on suggesting this may be caused by the perigalactic
passage of the Clouds and the Galaxy's gravitational attraction. This
coincides with the second age maximum in the SMC at about 220~Myr.

The age distributions of Cepheids in the OGLE-III data analyzed by
Subramanian and Subramaniam (2015) and in our OGLE-IV sample are
consistent. We observe a very similar age distribution with two peaks and
the age tomographies are also alike (see Fig.~14 and Fig.~10 in Subramanian
and Subramaniam 2015). The analysis of SMC CCs by Joshi \etal (2016) showed
a SF peak at $250\pm50$~Myr which is consistent with our older Cepheid SF
peak in this galaxy. They have also detected a second peak at about 160~Myr
in the eastern part of the SMC which is consistent with our conclusion from
Fig.~15, that the eastern part of this galaxy is younger.

\Section{Conclusions}
In this work we analyzed a total sample of 9418 fundamental-mode and
first-overtone CCs in the Magellanic System from the OGLE Collection of
Classical Cepheids based on the OGLE-IV data (Udalski \etal 2015, Soszyñski
\etal 2015). We fitted the P-L relations to the data using the Wesenheit
index for the {\it I}- and {\it V}-band photometry.  Fundamental-mode
Cepheids with $\log P\leq0.4$ were treated separately. The best fits for
the Wesenheit, the {\it I}- and {\it V}-band magnitudes are presented, for
both the LMC and SMC.

We calculated relative distances to each Cepheid using the reddening-free
Wesenheit index and the most accurate measurement of the mean LMC distance
from Pietrzyñski \etal (2013) as a reference. The results are presented on
three-dimensio\-nal maps in the Hammer equal-area projection and in the
Cartesian space.

The Cepheids in the LMC are present mainly in the bar and the northern arm.
Both structures, as well as the whole galaxy, are inclined such that the
eastern parts are closer to us. We fitted a plane to the LMC sample and
obtained the inclination and position angles of
$i=24\zdot\arcd2\pm0\zdot\arcd6$ and ${\rm
  P.A.}=151\zdot\arcd4\pm1\zdot\arcd5$ that are consistent with the
literature. The {\it rms} of our sample is $1.5$ kpc and it reflects the
significant scatter of the sample along the line of sight.

The age distribution of the LMC Cepheids reveals one peak at about 100~Myr.
Younger Cepheids tend to be clumped in the bar and the northern arm, while
older stars are spread all over the LMC disk. The northern arm seems to be
younger than the bar that has a similar age distribution as the whole
galaxy.

We redefined the LMC bar such that it spans almost the whole width of the
LMC. Both the classical bar (the central and eastern part of our bar) and
the newly added western part form one coherent structure that is clearly
visible in Cepheid density contours. Although the western part of the bar
is less numerous the two parts are connected both in their distance and age
on-sky distributions. Moreover, after the redefinition of the bar the
dynamical center of the LMC is now located in the center of the bar.

We separately fitted a plane to the bar Cepheids, despite the fact that this
may not be a proper physical model of the bar, although should yield a
reasonable offset.  The offset for the new bar is consistent with that for
the whole galaxy which means that the bar is not located closer to us than
the galaxy. On the other hand the distance distributions show that the
``classical'' bar that we call the eastern bar also is not offset from the
LMC plane, contrary to previous studies.

The LMC northern spiral arm is a very prominent feature in the Cepheid
distribution. We fitted a plane to the northern arm and found that this
structure is offset from the whole LMC sample by about 0.5~kpc toward us,
and lies in a different plane described by $i=34\zdot\arcd4\pm2\zdot\arcd9$
and ${\rm P.A.}=123\zdot\arcd8\pm3\zdot\arcd8$.

Our data does not reveal any other spiral arms in the central or southern
parts of this galaxy although we do see an additional spiral arm in the
north. We suppose that there may be another arm connected with the bar on
its south-eastern side, but there are too few Cepheids in that region to
provide strong evidence.

The unusual elongation of the SMC is confirmed in this study. The SMC is
elongated almost along the line of sight and its longitudinal dimension is
4--5 times greater than the transverse dimension. The north-eastern part of
the SMC is located closer to us than its south-western part. Note that both
Clouds are inclined toward each other.

The age distribution of the SMC Cepheids reveals two peaks, one at about
100~Myr, which is very similar to the LMC peak, suggesting a common star
formation epi\-sode, that could be due to LMC-SMC interaction, and another
one at about 220~Myr. Moreover, younger and older Cepheids are differently
distributed, supporting this hypothesis -- the former group is located in
the closer part of this galaxy, while the latter -- in the farther.

The SMC shape may be described as an extended ellipsoid with two additional
prominent off-axis structures that are also ellipsoidal. One is located in
the north of the SMC and is closer that the SMC main body and significantly
younger than the other one, which is located in the south-western part of
the SMC and hence farther.

The Wing of the SMC is not reflected in the Cepheid distribution, although
there are stars spread all over the galaxy and some of them in the eastern
part belonging to the Wing. Moreover, we see Cepheids at very large
distances ($\approx80$~kpc), that may be a stellar counterpart to the
Counter Bridge that is mixed with the SMC population.

The on-sky locations of most of the nine Magellanic Bridge Cepheids are
correlated with the young stellar population density contours. Moreover,
they seem to form a connection between the LMC and SMC. On the other hand,
the three-dimensional distribution of the Bridge CCs reveals that four of the
nine objects are located far from this connection, at very diverse
distances -- the closest one being closer to us than any of the LMC
objects, and the farthest one farther than any SMC Cepheid. This is an
important constraint for models of the Magellanic Clouds interactions.

All Bridge Cepheids except one have ages $<300$~Myr which is consistent
with the time of MBR formation and indicates that these stars were born
{\it in-situ}. The oldest MBR Cepheid may be connected with the SMC Wing
because of its nearby location.

\Acknow{We would like to thank Profs. M. Kubiak and G. Pietrzyñ\-ski,
  former members of the OGLE team, for their contribution to the collection
  of the OGLE photometric data over the past years. We thank the anonymous
  referee for comments and suggestions that greatly improved this
  publication.  A.M.J.-D. is supported by the Polish Ministry of Science
  and Higher Education under ``Diamond Grant''
  No. 0148/DIA/2014/43. D.M.S. is supported by the Polish National Science
  Center under the grant no.~2013/11/D/ST9/03445. I.S. is supported by the
  Polish Ministry of Science and Higher Education through the program
  ``Ideas Plus'' award No. IdP2012 000162. The OGLE project has received
  funding from the National Science Center, Poland, grant MAESTRO
  2014/14/A/ST9/00121 to AU.}

\end{document}